\documentclass[12pt,preprint]{aastex}

\newcommand{\kms}{\,{\rm km}\,{\rm s}^{-1}}
\newcommand{\hubunits}{\,{\rm km}\,{\rm s}^{-1}\,{\rm Mpc}^{-1}}
\newcommand{\halpha}{H$\alpha$}
\newcommand{\Mr}{M_r}
\newcommand{\Mg}{M_g}
\newcommand{\Mi}{M_i}

\newcommand{\Veight}{V_{\rm 80}}
\newcommand{\Vcom}{V_0}
\newcommand{\Vcirc}{V_{\rm circ}}
\newcommand{\Vend}{V_{\rm end}}

\newcommand{\Rend}{R_{\rm end}}

\newcommand{\Vflat}{V_{\rm flat}}

\newcommand{\flaga}{flag-1}
\newcommand{\flagb}{flag-2}
\newcommand{\flagc}{flag-3}
\newcommand{\flagd}{flag-4}
\newcommand{\flagbc}{flag-2,3}
\newcommand{\flagab}{flag-1,2}
\newcommand{\flagabc}{flag-1,2,3}
\newcommand{\flagabcd}{flag-1,2,3,4}
\newcommand{\flagcd}{flag-3,4}

\begin{document}

\title{The Tully-Fisher Relation and Its Residuals for
a Broadly Selected Sample of Galaxies}

\author{James Pizagno\altaffilmark{1}, 
Francisco Prada\altaffilmark{2,3},
David H. Weinberg\altaffilmark{1},
Hans-Walter Rix\altaffilmark{2},
Richard W. Pogge\altaffilmark{1},
Eva K. Grebel\altaffilmark{2,4},
Daniel Harbeck\altaffilmark{2,5}, 
Michael Blanton\altaffilmark{6},
J. Brinkmann\altaffilmark{7},
James E. Gunn\altaffilmark{8}
}

\altaffiltext{1}{Department of Astronomy, Ohio State University,
Columbus, OH 43210, USA}
\altaffiltext{2}{Max-Planck-Institut f\"{u}r Astronomie, K\"{o}nigstuhl 17, 
D-69117 Heidelberg, Germany}
\altaffiltext{3}{Ram\'{o}n y Cajal Fellow, Instituto de Astrof\'{i}sica de Andalucía (CSIC), E-18008 Granada, Spain}
\altaffiltext{4}{Astronomisches Institut, Universit\"{a}t Basel, 
CH-4102 Binningen, Switzerland}
\altaffiltext{5}{Department of Astronomy, University of Wisconsin, 
Madison, WI 53706, USA}
\altaffiltext{6}{New York University, Center for Cosmology and Particle
Physics, 4 Washington Place, New York, NY 10003}
\altaffiltext{7}{Apache Point Abservatory, P.O. Box 59, Sunspot, NM 88349}
\altaffiltext{8}{Princeton University Observatory, Peyton Hall, Princeton NJ 08544-1001}


\begin{abstract}
We measure the relation between galaxy luminosity and disk 
circular velocity (the Tully-Fisher, or TF, relation), in the 
$g$, $r$, $i$, and $z$-bands, for a broadly selected sample of 
galaxies from the Sloan Digital Sky Survey, with the goal 
of providing well defined observational constraints for 
theoretical models of galaxy formation.  The input sample of 
234 galaxies has a roughly flat distribution of absolute magnitudes 
in the range $-18.5$ $>$ $\Mr$ $>$ $-22$, and our only 
morphological selection is an isophotal axis-ratio cut 
$b/a$ $<$ 0.6 to allow accurate inclination corrections.  Long-slit 
spectroscopy from the Calar Alto and MDM observatories 
yields usable \halpha\ rotation curves for 162 galaxies (69\%),
with a representative color and morphology distribution.  We define
circular velocities $\Veight$\ by evaluating the rotation 
curve at the radius containing 80\% of the $i$-band light.  
Observational errors, including estimated distance errors due to peculiar 
velocities, are small compared to the intrinsic scatter of the 
TF relation.  The slope of the forward TF relation steepens from
$-5.5$ $\pm$ $0.2\,\hbox{mag}/\hbox{log}_{10}\kms$ in the $g$-band
to $-6.6$ $\pm$ $0.2\,\hbox{mag}/\hbox{log}_{10}\kms$ in the 
$z$-band.  The intrinsic scatter is $\sigma$ $\approx$ 0.4 mag in 
all bands, and residuals from either the forward or inverse relations 
have an approximately Gaussian distribution.  We discuss how
Malmquist-type biases may affect the observed slope, intercept, 
and scatter.  The scatter is not 
dominated by rare outliers or by any particular class of galaxies, 
though it drops slightly, to $\sigma$ $\approx$ 0.36 mag, if we 
restrict the sample to nearly bulgeless systems.  Correlations
of TF residuals with other galaxy properties are weak:  bluer 
galaxies are significantly brighter than average in the $g$-band
TF relation but only marginally brighter in the $i$-band; more 
concentrated (earlier type) galaxies are slightly fainter than 
average; and the TF residual is virtually independent 
of half-light radius, contrary to the trend expected for
gravitationally dominant disks.  The observed residual correlations 
do not account for most of the intrinsic scatter, implying that 
this scatter is instead driven largely by variations in the ratio of dark to 
luminous matter within the disk galaxy population.  
\end{abstract}


\section{Introduction}
The observed correlation between luminosity and disk 
rotation speed \citep{tully77} is one of the fundamental 
empirical clues to the physics of galaxy formation, in 
particular to the relation between dark matter halos and 
their luminous baryonic components.
The Tully-Fisher (hereafter TF) relation has been widely 
exploited as a distance indicator, in studies of the cosmic 
distance scale (e.g., \citealt{tully77,aar86,tul00,fre01}) 
and the large scale peculiar velocity field 
(e.g., \citealt{wil90,mat92,wil97b,cou00}).  The ambitious 
surveys constructed for such studies have usually focused on 
a relatively narrow range of galaxy types, typically 
undisturbed late-type spirals, with the goal of obtaining 
a tight relation that can yield precise distances.  
However, the small scatter and the measured parameters of the 
TF relation are adopted as key constraints on galaxy 
formation theories (e.g., \citealt{kau93,col94,mo98,som99,nav00}),
 even though these may not yet have the detail to predict the 
precise types of model galaxies.  The goal of this paper 
is to measure the TF relation for a broadly selected sample 
of galaxies, with a focus on quantifying rather than minimizing 
the intrinsic scatter and on measuring the 
correlation of TF residuals with other galaxy properties.

Our sample of galaxies is drawn from the Sloan Digital Sky 
Survey (SDSS, \citealt{yor00}), an imaging and spectroscopic 
survey of the North Galactic Cap and selected regions of the 
South Galactic Cap.  The SDSS galaxy spectra are obtained 
through a $3''$ diameter fiber, so they do not yield reliable
estimates of rotation velocities for galaxies that are 
spatially well resolved.  We have therefore obtained long-slit
 \halpha\ rotation curve data for a sample of 234  
SDSS galaxies using the Calar Alto 3.5-m telescope and the
 MDM 2.4-m telescope.  The Calar Alto observations of 189 
galaxies constituted a substantial portion of the Calar Alto 
Key Project contribution to the SDSS.  Selection from the 
SDSS redshift survey allows us to define our sample based 
on absolute magnitude rather than apparent magnitude, while 
working in a redshift range $5000\kms \leq v \leq 15000\kms$
 where peculiar velocities add little uncertainty to 
individual galaxy distances.  Equally important, the SDSS 
provides high quality, 5-band imaging for all of the program 
galaxies, and the connection to the much larger SDSS database 
allows statistical application of our results to, e.g., infer 
the distributions of galaxy potential well depths or angular 
momenta in the local universe.

As discussed in \S 2, our selection criteria 
produce a roughly flat distribution in $r$-band absolute 
magnitude over the range $-18.5$ $>$ $\Mr$ $>$ $-22$.\footnote{Throughout the
paper, we adopt $h$ $\equiv$ $H_0$ / 100
$\kms$ Mpc$^{-1}$ $=$ 0.7 and quote absolute magnitudes for this 
value.  One should add 5\,log($h$/0.7) for other values of 
$h$.  All logarithms are base 10.}  We select galaxies with isophotal axis ratios
$b/a \leq 0.6$ so that we can make accurate inclination corrections 
to rotation speeds, but we impose no other morphological selection 
criteria.  We obtain usable \halpha\ rotation curves for 170 of 
our 234 selected targets (73\%).  The axis ratio cut slightly 
reduces the representation of early-type galaxies, and the 
early-type galaxies that pass this cut are slightly less 
likely to yield usable \halpha\ rotation curves.  Nonetheless,
 relative to most previous large TF samples, our sample is 
more representative of the range of disk galaxy types.  In 
addition, the redshift range and accurate photometry 
make our typical observational errors smaller than 
the intrinsic TF scatter, which is essential for an accurate 
estimate of this scatter and a full understanding of residual 
correlations.  In brief, this is a TF sample designed 
for studies of galaxy formation, not for measurements of the 
Hubble constant or peculiar velocities.  We
presented and discussed the scaling relations for a disk-dominated 
subset of this sample in (\citealt{piz05}, hereafter P05).
The selection of inclined galaxies ensures a small uncertainty 
on the inclination corrected velocity width.  

There are a variety of ways to define a galaxy's rotation speed 
from optical or 21cm data, and in general they yield similar but
 not identical results for the correlation between luminosity 
and rotation speed (see  \cite{ver01} for 
a careful investigation of this issue).  As discussed in 
\S 4, our measure of disk rotation speed is based on 
the value of an arc-tangent fit to the rotation curve 
(\citealt{cou97}, hereafter C97) evaluated at a position containing 80\% of 
the total $i$-band flux.  The work of \cite{tully77} used 
21cm line widths rather than optical rotation curves, so 
strictly speaking our analysis is not the ``Tully-Fisher 
relation'', but we will follow common practice in using 
this term for the more general correlation between 
luminosity and gas rotation speed.

The observational study most similar to our own, in terms
of a broad sample selection, is that of \cite{kan02}, 
hereafter K02, who measured TF relations and residual 
correlations in $U$, $B$, and $R$-bands for a sample of 68 
galaxies with morphological types Sa-Sd from the Nearby
Field Galaxy Sample \citep{jan00}.  Our sample is larger by a 
factor of $\sim$ 2.5, and the luminosity distribution 
is rather different:  its absolute magnitude 
histogram is roughly flat in the range $-18.5$ $>$ $\Mr$ $>$ $-22$, 
with somewhat higher representation on the bright end, 
while K02's histogram peaks at $M_B$ $=$ $-18$ and declines at brighter
magnitudes.  The selection of our sample at $v$ $\geq$ 5000 $\kms$ 
and the use of SDSS surface photometry to determine disk 
inclinations makes our observational errors much smaller 
than those of K02, giving us a better handle on the TF intrinsic 
scatter.  However, unlike K02 we do not have literature 
HI data for many galaxies in our sample, nor do we extend 
our sample to faint luminosities.  In comparisons 
to the voluminous observational TF literature, we will mostly focus
on K02 and on the work of C97, who 
analyzed large samples of disk galaxy optical 
rotation curves using methods similar to those adopted here, 
and Verheijen (2001, hereafter V01), who carried out a detailed 
investigation of the TF relation in the Ursa Major cluster using 
resolved HI rotation curves and optical and infrared surface 
photometry.  

Numerous papers have used TF results to derive conclusions about 
the physics of galaxy formation and the relative importance of 
baryons and dark matter in the luminous regions of disk galaxies 
(e.g., \citealt{fab82,gun82,per88,col89,whi91,kau93,col94,
eis96,per96,dal97,mo98,cou99,fir00,nav00,dut06,gne06}).  
Our underlying goal is to provide better 
observational inputs for these kinds of theoretical and modeling 
efforts.  We have therefore paid particular attention to 
characterizing our sample selection and observed quantities as 
straightforwardly as we can, and to estimating the intrinsic scatter 
in the different SDSS bands.  We have already presented the 
scaling relations and inferred baryonic mass 
fractions of a disk-dominated
subset of our full sample in P05, and \cite{gne06} have 
used these results to constrain the distributions of disk-to-halo 
mass ratios and spin parameters of spiral galaxies.  

\section{SDSS Observations and Sample Selection}
The SDSS uses a mosaic CCD camera \citep{gun98} on a dedicated 
2.5-m telescope \citep{gun06} to
image the sky in five photometric band passes 
\citep{fuk96} denoted $u$, $g$, $r$, $i$, 
$z$.\footnote{\cite{fuk96} actually define a slightly 
different system, denoted $u'$, $g'$, $r'$, $i'$, $z'$, 
but SDSS magnitudes are now referred to the native 
filter system of the 2.5-m survey telescope, for 
which the bandpass notation is unprimed.}  The 
imaging data are reduced by a series of automated 
pipelines that perform astrometric calibration 
\citep{pie03}, photometric data reduction \citep{lup01,sto02}, 
and photometric calibration \citep{fuk96,hog01,smi02,ive04,tuc06}.  
We choose our TF sample from the main galaxy spectroscopic sample, 
which is selected from the SDSS imaging data using 
the algorithm described by \cite{str02}.  

As discussed in the introduction, our goal is to investigate 
a galaxy sample that is as representative as possible of the 
full galaxy population, while keeping our sample completeness
 high and our observational errors significantly below the 
expected intrinsic scatter of the TF relation.  We have 
chosen to focus on the absolute magnitude range $-18.5 > \Mr > -22$. 
 Brighter galaxies tend to be early-type systems for which 
it is difficult to obtain \halpha\ rotation curves, and the
 behavior of the TF relation at fainter magnitudes, while an
 interesting problem in itself, brings in additional 
complications because of the morphological irregularity 
and low rotation speeds of the galaxies.  

We correct our observed velocities for inclination using 
observed disk axis ratios, but select galaxies in such a way that any 
possible intrinsic disk ellipticities contribute a negligible amount 
of uncertainty to the correction.  The intrinsic ellipticities of disks, thought to be 
$\sim$ $5-10\%$ rms  depending on galaxy 
type and band \citep{zar95,ryd05}, are an irreducible source 
of uncertainty in TF studies without 2D velocity fields, 
since they prevent one from 
perfectly measuring a galaxy's inclination and thus inferring 
its deprojected rotation speed.  We have selected galaxies to 
have a measured axis ratio $b/a$ $\leq$ 0.6, so that a 5\% 
intrinsic ellipticity would change the inclination corrected
 rotation velocity 
$V_{\rm obs}/\sin i$ $\propto$ $V_{\rm obs}(1-b^2/a^2)^{-1/2}$ by $\leq 5\%$.  
Random ellipticities of this order would then add $\sim 0.15$ 
magnitudes of scatter to the forward TF relation for a typical
slope $L \sim V^3$, which is smaller than typical estimates 
of the intrinsic scatter by a factor $\sim 2-3$.  Note, however, 
that disk ellipticities also contribute scatter via 
non-circular motions; see the discussion in \S 5.3 below.    
Specifically, we base our sample selection on the SDSS $r$-band 
isophotal axis ratio ({\tt isoB\_r/isoA\_r}), which is measured 
at an isophote of 25 mag$\,{\rm arcsec}^{-2}$ where the disk 
typically dominates over the bulge.  We ultimately base our 
inclination corrections on the results of our 2-D bulge-disk 
decompositions described in \S 4.3 below.  The intrinsic 
ellipticity of a disk is not the same as the intrinsic axis ratio, 
where the intrinsic axis ratio is the observed axis ratio when a galaxy 
is viewed edge-on (see \S 4.5 equation 3).  

We assume that typical galaxy peculiar velocities relative to the local
large scale flow, $\sim$~200$-$300$\kms$ \citep{str95}, 
introduce only a small uncertainty in galaxy luminosities.  We therefore 
select our sample to have galaxy redshifts $cz > 5000\kms$,
 so that the corresponding absolute magnitude uncertainties 
are $\Delta M \la 5\log(1+300/5000)=0.13$ mag.
At the same time, we want galaxies to be at least several 
arc-seconds across, so that we can get good morphological 
measurements from the images and $\sim$2$''$ seeing does 
not seriously degrade rotation curve measurements.  Since 
galaxy intrinsic sizes correlate with luminosity, we 
therefore impose a maximum redshift that depends 
on absolute magnitude:  $9000\kms$ for $-18.5 \geq \Mr > -20$, 
$11,000\kms$ for $-20 \geq \Mr > -21$, 
and $15,000\kms$ for $\Mr<-21$.\footnote{These cuts are not exact 
because of small changes in the SDSS photometry and our 
selection criteria over the course of the survey, but they are a
close approximation to our redshift limits.}  
These cuts ensure that all galaxies
have a half-light diameter $2r_{50} > 4''$; the median 
value in the sample is  $2 r_{50} = 19.5 ''$, and the 10\% and 
90\% values are 11.3$''$ and 34.8$''$ (based on the 
$i$-band bulge-disk decomposition described in 
\S 4.3 below).  The faintest apparent magnitude allowed 
by these cuts (for $\Mr$ = $-18$ galaxies at 9000 $\kms$) is 
$r = 17.5$, brighter than the SDSS main galaxy spectroscopic 
limit $r$=17.7 mag.  The 
median apparent magnitude of the sample is $r = 15.06$ mag.
  We compute galaxy luminosities using SDSS Petrosian fluxes and 
colors using SDSS model colors, both $K$-corrected to 
redshift $z=0$  using Blanton et al.'s (\citeyear{bla03a})
{\tt kcorrect\_v3.1b}.  We compute distances using the 
SDSS heliocentric redshifts corrected to the rest frame 
of the Local Group barycenter (\citealt{wil97a}),
assuming a cosmological model with $\Omega_m$=0.3, 
$\Omega_{\lambda}$=0.7, and $h$ = 0.7. 
 We incorporate a distance uncertainty corresponding to $300 \kms$ 
when calculating disk scale length and luminosity uncertainties, 
to account for the typical amplitude of small scale peculiar 
velocities (\citealt{str95}).

This combination of absolute magnitude and redshift limits 
gives a distribution of candidates that is roughly flat in 
absolute magnitude over the range $-18.5$ $>$ $\Mr$ $>$ $-22$,
since we search for rarer, brighter galaxies over a larger 
volume.  We do not impose any explicit cut at $M_r=-22$ mag, 
but we have relatively few brighter galaxies in our sample 
because of the declining luminosity function.
Ideally, we would search for all galaxies in a relatively 
narrow shell, e.g. $5000 \kms < cz < 7000 \kms$, so that 
they would be as well resolved as possible, and we would
 use weighted random sampling to obtain a flat 
distribution in absolute magnitude.  However, we began 
our observations in June 2001 when the sky area covered 
by the SDSS was still relatively small, and we needed to 
go to larger distances for brighter galaxies to obtain a 
sufficient number of targets.  We have kept our selection 
criteria fixed throughout the course of the observing program, 
rather than lower the outer redshift limits as the SDSS 
sky coverage increased.  All of the galaxies in our sample 
are included in the SDSS Second Data Release (DR2; \citealt{aba04}),
 and when possible we have taken images and photometric
 parameters for the final analysis in this paper from the
 public database.  The selection of candidates was based 
on the state of reduction of the imaging data available 
at the time of our observations, so there might be small 
differences between the parameters used for selection 
and the parameters used for analysis. The only SDSS 
spectroscopic parameter that we use is the redshift,
 to determine galaxy distances.

Figure~\ref{fig:histogram} shows the $r$-band absolute magnitude 
distribution of our sample.  The open histogram shows 
all of the galaxies that we spectroscopically targeted, and the filled 
histogram shows those galaxies that yielded \halpha\ 
rotation curves good enough for use in our TF study.  
As discussed in \S 4.1 below, we characterize galaxies 
with unusable rotation curves as \flagd, those with 
rotation curves still rising at the outermost data point
 as \flagc, those with rotation curves just reaching 
the turnover region as \flagb, and those with extended
 flat portions of the rotation curve as \flaga.  Most 
\flagd\ galaxies simply lack a sufficient amount of extended 
\halpha\ emission, but in four cases they have an extended
 emission pattern that has too low signal-to-noise
\halpha\ emission to allow sensible definition of a rotation velocity.  
Overall, we obtained usable \halpha\ rotation curves 
(\flagabc) for 170 of our 234 target galaxies, or 73\%. 
 After the bulge-to-disk decomposition, discussed in
 \S 4.3, we compared our disk position angle 
to the SDSS isophotal position angle and removed six
galaxies for which the difference between the SDSS 
and our position angles would cause a $>$10\% 
difference in rotation velocity.  
We also removed two galaxies that have disk axis ratios
greater than 0.7, as derived from our bulge-disk decomposition. 
The final sample size is 162 galaxies.  The filled 
histogram is approximately flat in absolute magnitude,
but there are fewer low luminosity systems, in part 
because we concentrated our early observations on the
range $\Mr$~$<$~$-19.5$.  In our TF analysis, we will 
quote results both for the sample ``as is'' and for 
the sample weighted to approximate a truly flat 
absolute magnitude distribution over the range 
$-18.5$~$>$~$\Mr$~$>$~$-22$, since the latter provides a 
well defined target for theoretical models to 
predict.  In practice, this reweighting makes little 
difference to the results (see \S 5.2).

Ideally, we would like to obtain rotation velocities for a 
random subset of {\it all} galaxies in the absolute 
magnitude range $-18.5$~$>$~$\Mr$~$>$~$-22$.
Our only explicit morphological pre-selection is the 
axis ratio requirement $b/a<0.6$, which is necessary 
to allow accurate inclination corrections to the measured 
velocities.  Since disks are flatter than bulges, this axis
 ratio cut tends to suppress the representation of early 
type galaxies in the sample.  The requirement of obtaining 
a usable \halpha\ rotation curve imposes another, less well-controlled 
selection on galaxy type --- the galaxies in our 
final sample are those with a sufficient amount of extended, 
ionized gas.  In principle, we could obtain rotation curves 
for other galaxies using stellar absorption lines, but these
 require considerably longer exposures, and we did not have 
enough observing time to get a large sample and go deep enough for 
absorption line rotation curves.  Completing the sample 
with absorption line rotation curves would be a valuable 
follow up to the present study, requiring a comparable 
amount of observing time.  

\begin{figure}
\plotone{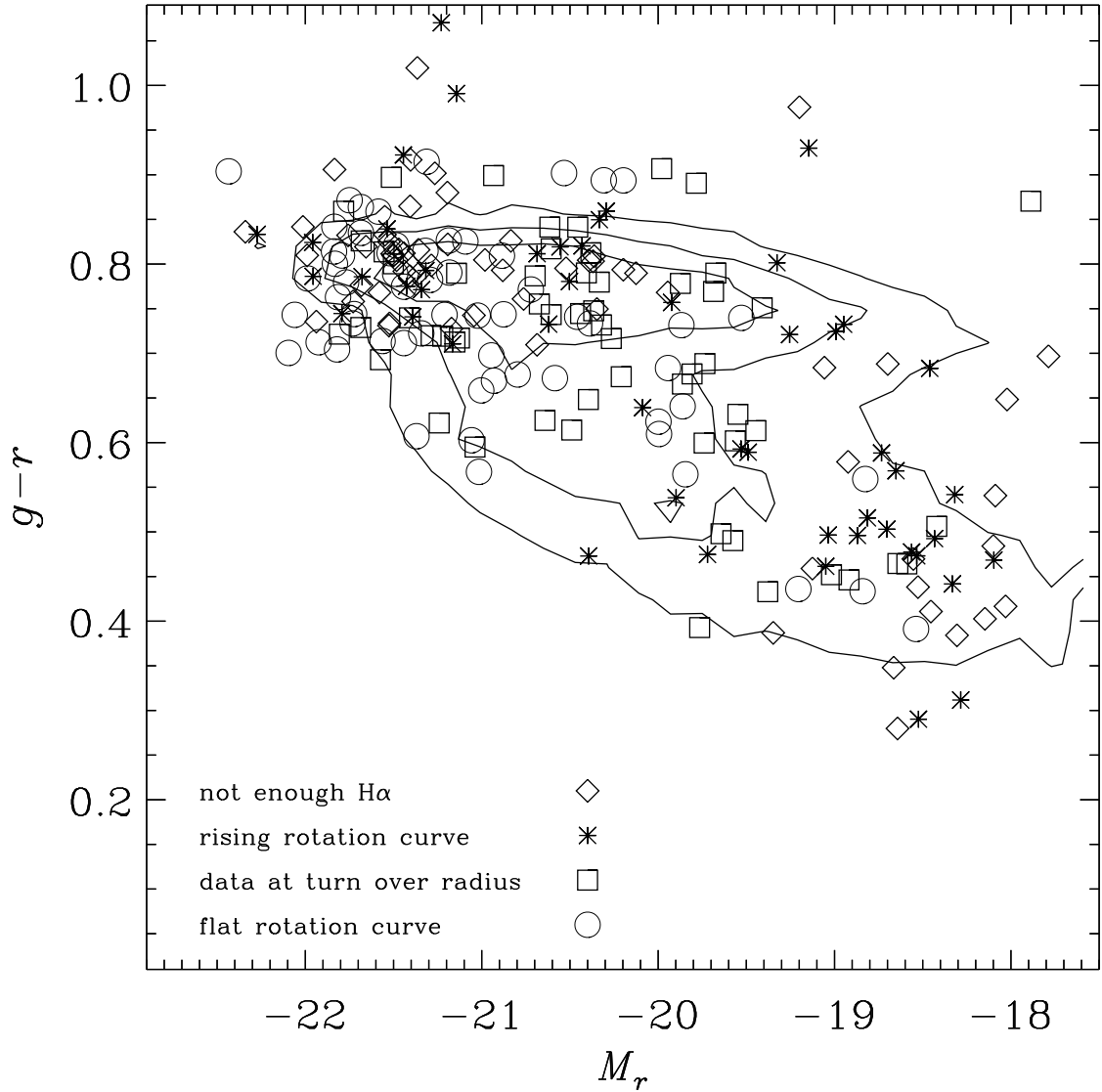}
\caption{The color magnitude relation for our sample of galaxies plotted 
on top of contours containing 25\%, 50\%, and 75\% of the total luminosity
density in the whole low redshift DR2 sample \citep{bla03b}.
The data point type indicates the quality of the galaxy rotation curve 
corresponding to \flaga\ (circles), \flagb\ (squares), 
\flagc\ (asterisks), and \flagd\ (diamonds).}
\label{fig:colormag}
\end{figure}

Figure~\ref{fig:colormag} shows the distributions of our sample galaxies 
in the color-magnitude plane, $\Mr$ vs. $g$-$r$.  Contours 
show the luminosity-weighted 
color magnitude distribution for the full DR2 sample of galaxies 
from \cite{bla03b}, with contours containing 25\%, 50\%, and 
75\% of the DR2 total luminosity density.   The four 
data point types indicate the quality of the rotation curve obtained 
for the galaxies.  Figure~\ref{fig:colormag} shows the distribution of our galaxies relative 
to a sample without velocity measurements.  
Galaxies without extended \halpha\ emission are concentrated at 
low luminosity, where the surface 
brightness is low, and at high-luminosity and red colors, where 
galaxies are more likely to be gas poor.  Galaxies with 
rising rotation curves are also more concentrated at low luminosity.  
However, there are usable data in all 
regions of the color-magnitude plane.  In accord with the 
usual trends for luminosity-environment correlations 
(e.g. \citealt{hog04}), the high luminosity galaxies 
are predominantly in over-dense regions, while the lower luminosity 
galaxies are found in less dense regions.  

\begin{figure}
\epsscale{1.0}
\plotone{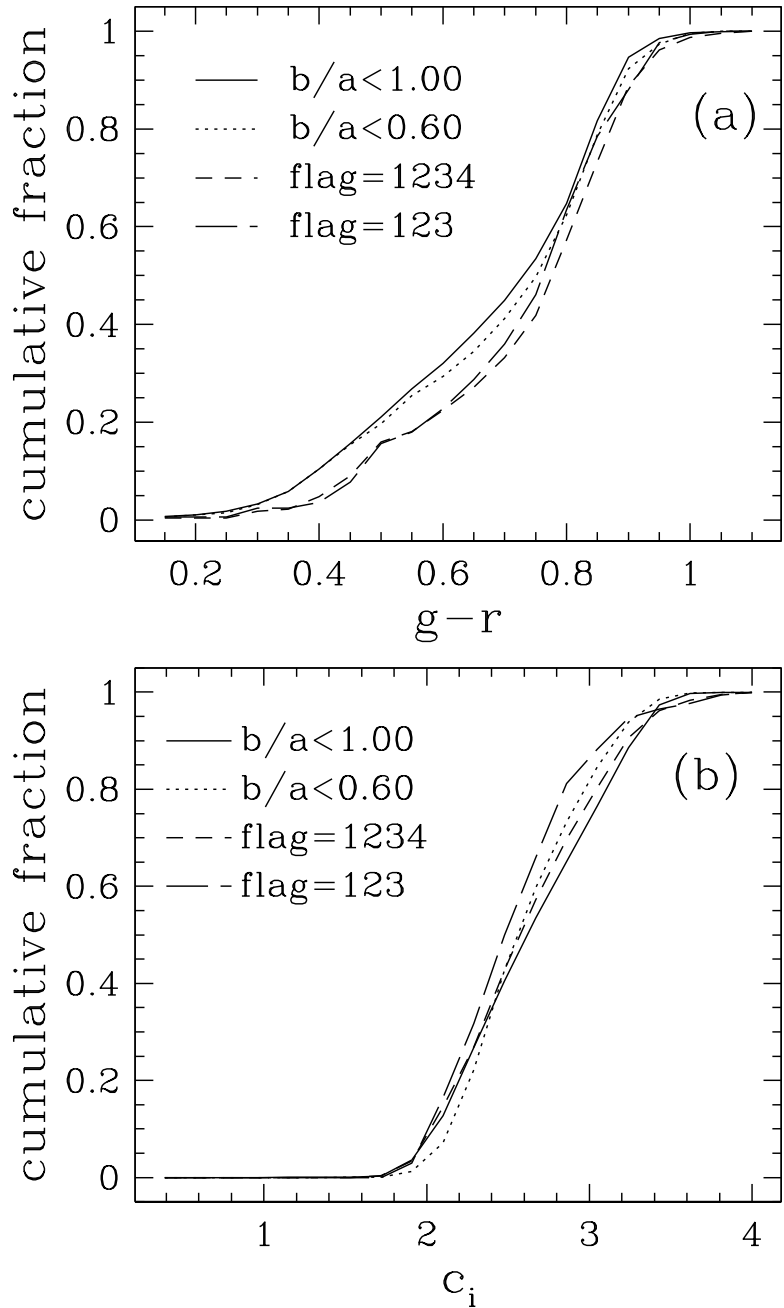}
\caption{ Cumulative distributions of the (top) $g$-$r$ model color and 
(bottom) $i$-band concentration index ($r_{90}/r_{50}$) for our full sample 
(short-dashed curves), the subset with usable rotation curves 
(long-dashed curves), and ``control" samples with the same $\Mr$ 
distribution selected from SDSS DR4, with (dotted) and without (solid) 
an isophotal axis ratio cut $b/a$ $\leq$ 0.6.}
\label{fig:cumdist}
\end{figure}

Figure~\ref{fig:cumdist} compares the properties of our sample to a ``control''
sample with the same $\Mr$ distribution from SDSS 
DR4 \citep{ade06}.  We have selected a large set of galaxies 
from DR4,\footnote{Using the Skyserver website for DR4:  
{\tt http://skyserver2.fnal.gov/dr4/en/}.} 
that matches our full sample's $\Mr$ distribution 
(open histogram of Figure~\ref{fig:histogram}) but has 22 times more 
galaxies in each 0.5 magnitude bin, thus minimizing statistical 
fluctuations in the control sample.  The cumulative $g$-$r$ 
distribution of the full DR4 sample is shown 
as a solid line in Figure~\ref{fig:cumdist}a.  The dotted curve shows the 
distribution after restricting this sample to galaxies with $i$-band 
isophotal axis ratio $b/a$ $\leq$ 0.6; the impact of the axis 
ratio cut is tiny.  The short-dashed and long-dashed lines 
show our \flagabcd\ and \flagabc\ distributions.
 Comparing the short-dashed and long-dashed lines, we find 
that selecting galaxies according t0 \halpha\ detectability 
does not alter the color distribution.  There is a 
small but statistically significant difference between our 
sample of (\flagabcd) and the DR4 sample with $b/a$ $\leq$ 0.6.  We have been 
unable to identify the cause of this difference, since in 
principle the two samples should differ only in size, but it is 
small in any case.

Figure~\ref{fig:cumdist}b shows the cumulative distributions of the $i$-band concentration 
index ($c_{i}$ = $r_{90}$/$r_{50}$), where $r_{90}$ and 
$r_{50}$ contain 90\% and 50\% of the Petrosian flux.  A typical ``early-type'' 
division is $c_{i}$ $\geq$ 2.6 \citep{str01}; see Figure~\ref{fig:dtihub} below.
Comparing the solid and dotted lines in Figure~\ref{fig:cumdist}b shows that 
the early-type fraction is reduced when the axis ratio 
cut is applied, but the effect is small.  Comparing 
the short-dashed to the long-dashed lines shows that 
selecting galaxies according to \halpha\ detectability 
reduces the early-type fraction, but again the effect is 
small.  We conclude that our axis ratio cut and 
requirement of \halpha\ detectability do not 
strongly alter the color or concentration distribution 
of the galaxy population 
when compared to a random sample of galaxies with the 
absolute magnitude distribution shown in Figure~\ref{fig:histogram}.  In 
this $\Mr$ range, the ``true'' early-types (i.e., ellipticals and 
bulge-dominated S0s) that would be strongly 
suppressed by the axis ratio cut are rare, and the high 
completeness of the \halpha\ observations leaves little 
room for further selection.  Classifying the \flagabcd\  
galaxies according to Hubble type, shows that 
the \flagd\ galaxies are predominantly Sa with a few, if any, 
true S0 galaxies. 

\section{Spectroscopic Observations and Data Reduction}
Spectroscopic observations were carried out at the Calar Alto Observatory
using the TWIN spectrograph mounted on the 
3.5-m telescope, and at the  MDM Observatory using the CCDS 
spectrograph mounted on the 2.4-m Hiltner 
telescope.  Observations at Calar Alto were carried out between 
June 2001 and October 2002; 30 nights 
were allocated in total, but many were clouded out.  Observations 
at MDM were carried out between February 2003 
and April 2004, on a total of 18 nights.  A total of 237
spectra were taken, with 52 spectra from MDM and 185 from Calar Alto.
Spectra for 3 galaxies were repeated at Calar Alto and MDM 
to allow comparison between the two telescopes 
(see the end of this section).  As discussed above and in \S 4.1,
170 of the 234 un-repeated spectra yielded usable \halpha\ rotation curves, 
and eight galaxies were eliminated from the sample based on position angle 
misalignments or axis ratios.  

The initial Calar Alto observations were 
carried out with total exposure times  
of 1800 seconds, chosen to provide a high 
signal-to-noise ratio on the \halpha\ line.  The observing 
strategy was updated in later observing runs 
to 1200 seconds for bright galaxies, 
1800 seconds for medium galaxies, and two 1200 second exposures for 
faint galaxies.  Most MDM observations took three 1200 
second exposures.  The last MDM run, on which 
we obtained spectra for 18 
galaxies, took an initial exposure for 1200 seconds.  Galaxies that 
showed no \halpha\ in the first exposure were not re-observed.  
If \halpha\ was detected in the initial exposure, then three more 
1200 second exposures were taken, making the total 
exposure time 4800 seconds.  Table~\ref{tbl:spectrographs} summarizes 
the main characteristics of both spectrographs and
the spectrograph setup.  The data 
were reduced using standard IRAF\footnote{IRAF is written 
and supported by the IRAF programming group at the National
 Optical Astronomy Observatories (NOAO) in Tucson, Arizona. 
NOAO is operated by the  Association of Universities for 
Research in Astronomy (AURA), Inc. under cooperative agreement
 with the National Science Foundation.} routines and the 
XVista\footnote{XVista can be found at:  
http://ganymede.nmsu.edu/holtz/xvista} package. 
 
The raw MDM and Calar Alto data were reduced in a similar 
manner. The data were corrected for bias and dark currents 
and flat fielded following procedures outlined in 
the Kitt Peak Low-to-Moderate Resolution Optical 
Spectroscopy Manual.\footnote{The manual can be 
found at:  http://www.noao.edu/kpno/manuals/l2mspect/spectroscopy.html}
Wavelength calibration and linearization were performed
using neon-arc lamps.  Cosmic rays were removed from the 
MDM data through median filtering of multiple images and
the IRAF routine .CRREJECT., or by hand for the Calar Alto 
data.  The telluric lines were also 
used to test the alignment and wavelength calibration of 
the spectra.

\begin{figure}
\epsscale{1.0}
\plotone{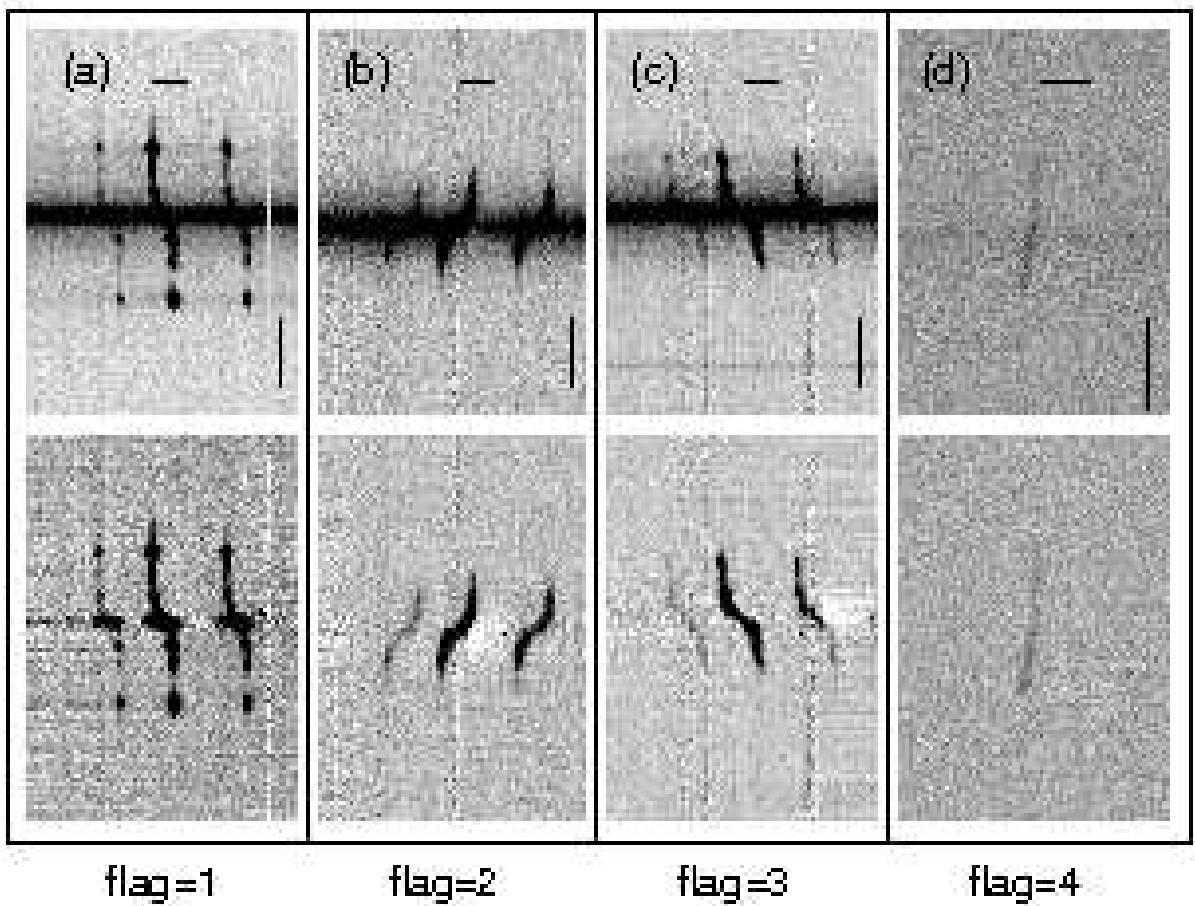}
\caption{ Examples of flat-fielded, wavelength 
calibrated, and linearized spectra of four galaxies
with varying quality flags for their \halpha\ rotation curves.  
Column (a) shows a \flaga\ spectrum, with data on the flat 
portion of the rotation curve.  Column (b) shows a 
\flagb\ spectrum, with data extending just past the 
turn-over radius.  Column (c) shows a \flagc\ spectrum, 
with a rising rotation curve at the last detectable 
data point.  Column (d) shows a \flagd\ spectrum, with 
insufficient \halpha\ for a useful rotation curve.  
The horizontal bars indicate 10 \AA\ and the vertical 
bars indicate 20 arc-seconds.  Lower panels show the spectra 
after continuum subtraction.  The central ``curve'' 
is the \halpha\ emission line used for the rotation curve
measurement; (a)-(c) also show the [NII] emission lines.}
\label{fig:spectra}
\end{figure}

Accurate measurements of velocity centroids at each 
spatial position along the slit require accurate 
removal of the galaxy continuum.  
A noise image was made that included the galaxy 
continuum to accurately account for the Poisson noise 
due to continuum plus \halpha\ photons.  The galaxy 
continuum was removed by averaging $\sim$10\AA\ of 
data on each side of the emission lines, and subtracting 
this from the area containing the \halpha\ emission.  
Telluric emission lines were subtracted from the data by averaging
 rows in the spatial direction, where there was no galaxy 
data, and subtracting that from the area of the CCD 
containing the galaxy data.  Tests using the wavelengths of 
telluric lines tabulated by \cite{ost96} show that the dispersion axis was 
aligned to be perpendicular to the columns of the CCD to 
an accuracy of 0.1 \AA.  Four flat-fielded, wavelength 
calibrated, and linearized spectra are shown in Figure~\ref{fig:spectra}, 
illustrating varying levels of \halpha\ detectability 
and spatial extent.     

\begin{figure}
\epsscale{1.0}
\plotone{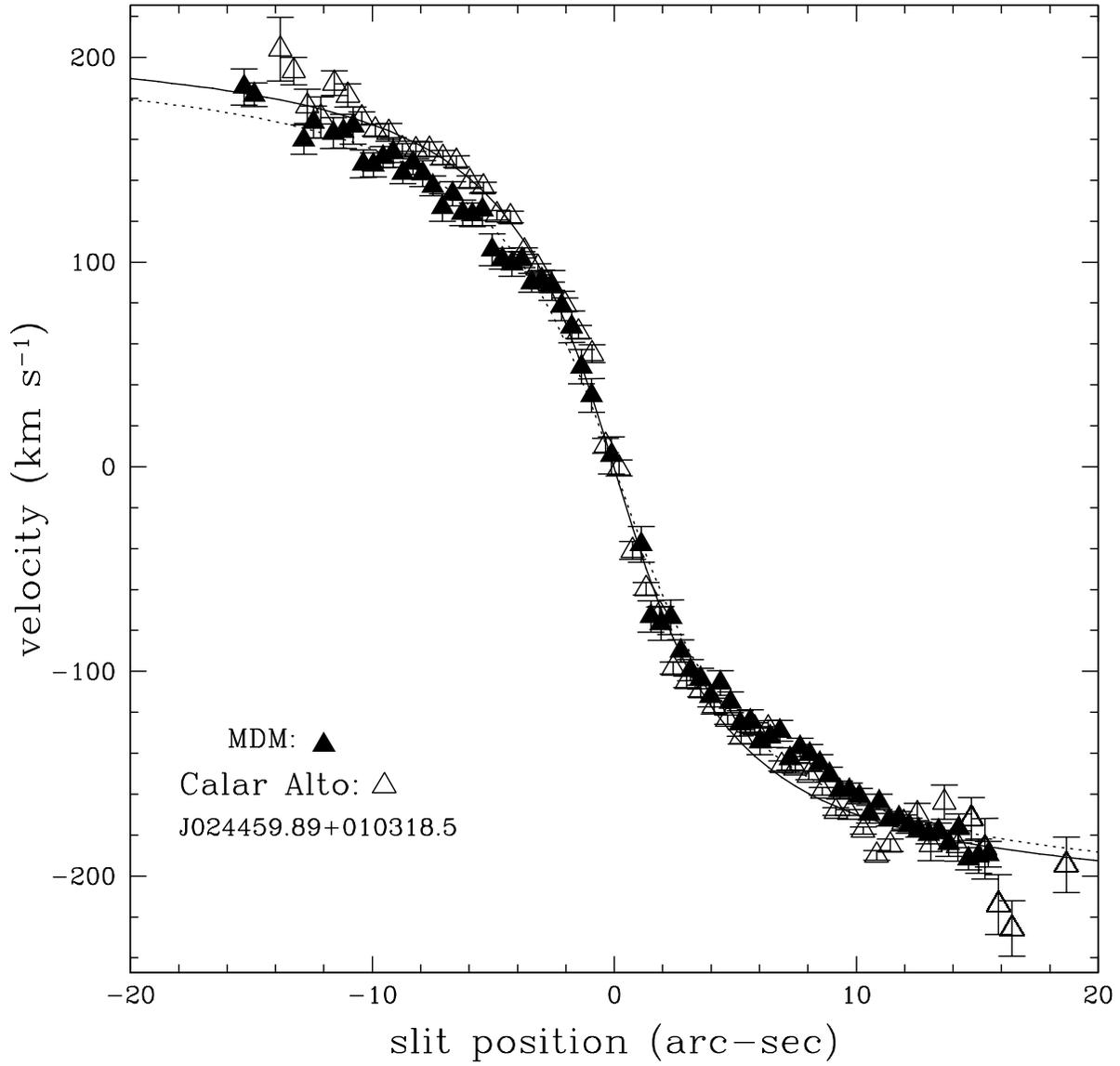}
\caption{Comparison of rotation 
curves measured at MDM (filled triangles) and Calar 
Alto (open triangles).  Dotted and solid curves show the 
arc-tangent function fits to the MDM and Calar Alto 
data, respectively.}
\label{fig:CAMDM}
\end{figure}

Figure~\ref{fig:CAMDM} compares the rotation curves for the galaxy SDSS  
J024459.89+010318.5, observed at both Calar Alto and MDM.  
The overall shapes of the rotation curves 
are similar, with slightly lower velocities in the MDM data.
The two arc-tangent function fits (see \S 4.1 below) are similar.   
The arc-tangent fit velocities at 18.4$''$ along the arc-tangent 
functions, the radius containing 80\% of the $i$-band flux 
(see \S 4.2), differ by 4.7 $\kms$, about 1.3$\sigma$ 
(MDM$=186.0\pm 3.3\,\kms$, Calar Alto$=190.7\pm 3.5\,\kms$).  
The other two galaxies with MDM and Calar Alto spectra 
show even better agreement in measured rotation speeds than 
the example shown in Figure~\ref{fig:CAMDM}.  We used the Calar Alto 
rotation curve for this galaxy, because the systematically
lower MDM rotation curve suggests a modest slit mis-alignment.
Regardless of these small differences there is still 
overall good agreement between the Calar Alto and MDM data.

\section{Measuring Rotation Velocities and Inclination-Corrected Luminosities}
\subsection{Rotation curve fitting}
The procedure to extract rotation curves from the spectra 
is similar to that used by C97.  The 2-D spectra for 
a galaxy are extracted into 1-D linear spectra,
 with each 1-D linear spectrum along the spatial 
direction of the slit being 1 pixel wide ($0.41 ''$ for MDM and 
$0.55 ''$ for Calar Alto).  
To avoid assuming an implicit shape of the emission 
lines, we measure the intensity weighted 
velocity centroid of the \halpha\ line in 
each 1-D spectrum.  The uncertainty in the \halpha\ 
centroid is measured using the signal-to-noise ratio (SNR) 
at each pixel with H$\alpha$ flux.    
The \halpha\ line centroid has typical
 uncertainties of $2-12$ $\kms$, depending on the 
total SNR of the emission line. 
 
Following C97, we fit the observed 
data points with an arc-tangent function,  
which has a minimal number of free parameters 
while adequately describing most galaxy 
rotation curves over the range probed by \halpha\ observations.  
Specifically, we fit the parameters $\Vcom$, $r_0$, $r_t$, and 
$\Vcirc$ of the relation
\begin{equation}
V(r_i)=\Vcom + \frac{2}{\pi}\Vcirc \mbox{ }{\rm arctan} \left (\frac{r_i-r_0}{r_t} \right)
\end{equation}
by minimizing 
$\chi^2$=$\sum_i$ $[$ $V(r_i)$-$V_{\rm obs}(r_i)$ $]^2$/$\sigma_i^2$, 
where $\Vcom$ is the systemic velocity, $r_0$ is the 
spatial center of the spectrum, $r_t$ is a 
turn-over radius where the rotation curve 
goes from steadily rising to flat, $\Vcirc$ 
is the asymptotic circular velocity, $V_{\rm obs}(r_i)$ is the
intensity weighted velocity centroid at pixel $r_i$, and $\sigma_i$ 
is its uncertainty.  Note that 
we do not force $r_0$ to correspond to the photometric
 center of the galaxy.  We use a Levenberg-Marquardt 
minimization routine \citep{pre92}, with initial guesses 
made by eye, to obtain the best-fit parameters and
 an error covariance matrix.  
Since galaxy disks have non-circular motions at the 
$10-20$~$\kms$ level and we do not want the fit dominated 
by the high SNR data points at the inner parts of the 
rotation curve, we add 10$\kms$ in quadrature to all 
the velocity centroid uncertainties.  Changing 10$\kms$ to 
20 $\kms$ makes a negligible difference in the best-fitting 
parameters.  Changing 10$\kms$ to 0$\kms$ makes little 
difference in most but not all cases.  The typical $\chi^2$/d.o.f. 
is $0.1-1.0$ when 10$\kms$ is added to the centroid uncertainty and
$1-2$ without this addition.    

Rotation curves are assigned flags indicating  
how well the data sample the flat part of the rotation 
curve.  Flag-1 indicates that there are data 
along the flat part of the rotation curve.  Flag-2 
indicates that there are data at the 
turn-over radius, but not beyond.  
Flag-3 indicates that an arc-tangent function 
adequately describes the data but the rotation 
curve is still rising at the last measured data 
points.  We assign \flagd\ to all galaxies with rotation 
curves either cannot be fit by an arc-tangent function, or 
have no detectable \halpha.  Although the flags 
are assigned by eye, the flag correlates well with 
the difference between the fit parameter $\Vcirc$ and the velocity ($\Vend$) at the 
last measured point of the rotation curve.  Out of 
234 galaxies, there are 64 \flagd, 50 \flagc, 57 \flagb, 
and 63 \flaga\ rotation curves.

\begin{figure}
\plotone{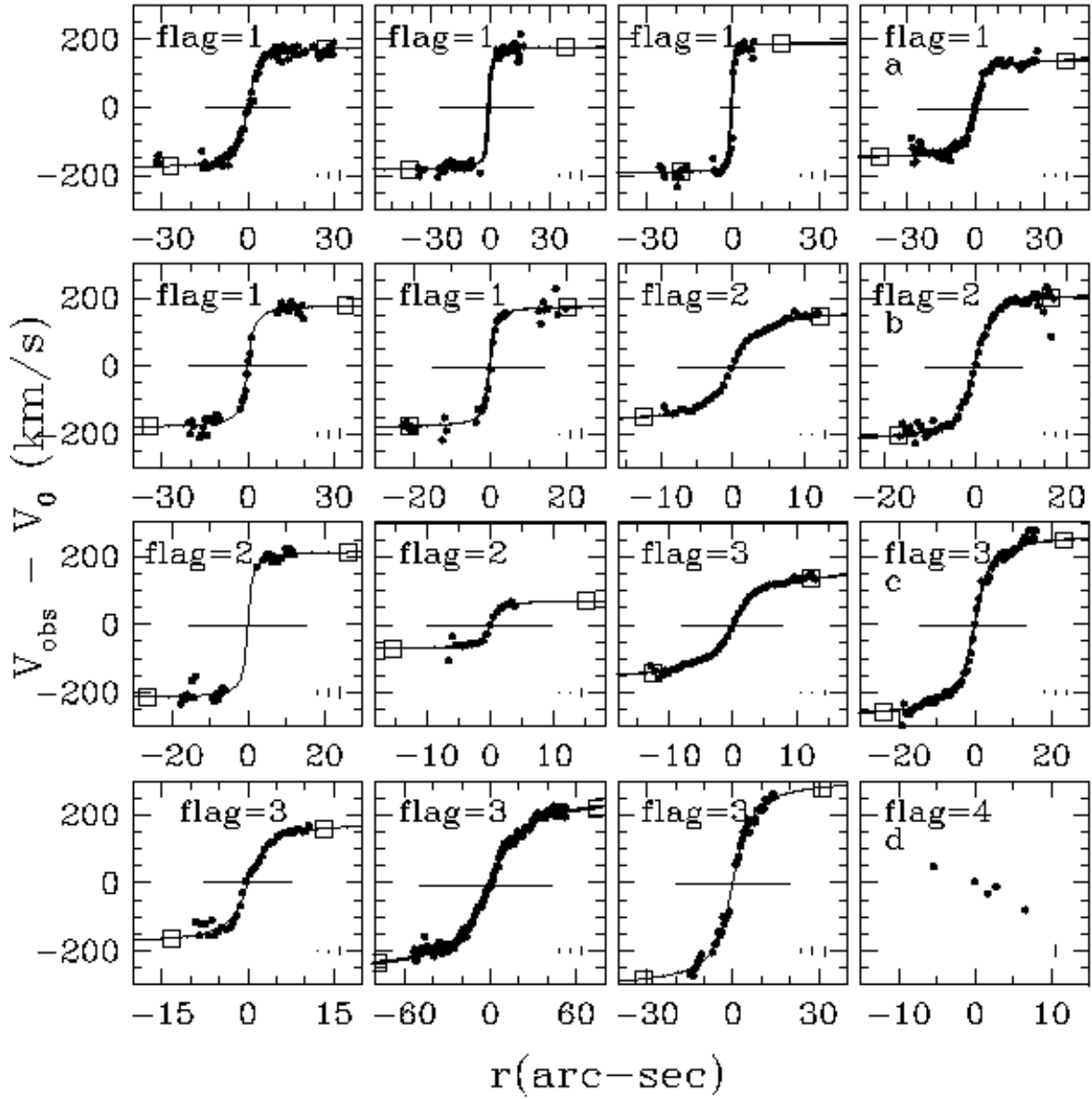}
\caption{Rotation curves of 16 sample galaxies.  
The data quality flag is marked in each panel, and 
the four galaxies shown in Figure~\ref{fig:spectra} appear in the 
right hand column.  
Vertical line segments in each panel 
show the 10th-percentile, median, and 90th-percentile 
values of the 1$\sigma$ error bars on the
observed velocity centroids.  Solid curves show the best-fit 
arc-tangent functions.  Open squares 
mark the fitted velocity at the radius enclosing 80\% 
of the $i$-band flux, which is our standard measure 
of the galaxy rotation speed.
Horizontal bars extend from $-2.2R_d$ to $+2.2R_d$, where
$R_d$ is the disk exponential scale length of the $i$-band
GALFIT model.
}
\label{fig:6090rot}
\end{figure}

Figure~\ref{fig:6090rot} shows rotation curves and best-fit arc-tangent functions for 16  
galaxies with flags 1 through 4; the four galaxies 
whose 2-D spectra appear in Figure~\ref{fig:spectra} are shown in the 
right-hand column.  Vertical line segments in each panel 
show the 10th-percentile, median, and 90th-percentile 
values of the 1$\sigma$ error bars on the
observed velocity centroids;  typical values are 4.5 $\kms$, 5.6 $\kms$, 
and 11.0 $\kms$.  The arc-tangent curves provide good descriptions 
of the overall shape of the rotation curve in all cases, 
but some rotation curves show ``bumps and wiggles'' associated 
with non-circular motions.  As discussed below, our primary 
measure of galaxy rotation speed is the value $\Veight$\ of the
arc-tangent fit at the radius containing 80\% of the $i$-band 
flux.  This point is marked by an open square in each panel. The 
horizontal lines have a full width of twice 2.2$R_d$, where $R_d$ is
the disk exponential scale length (see \S 4.3). 
 
\subsection{Rotation Speed Definitions}
For TF purposes, we want to characterize the 
fitted rotation curve by a single rotation 
velocity.  While the asymptotic circular speed 
$\Vcirc$ seems the obvious quantity to use 
when the rotation curve is fully constrained, the fitted 
value can vastly overestimate the true rotation speed 
when the observed data points do not reach the flat portion 
of the rotation curve.  The velocity $\Vend$ evaluated 
from the arc-tangent function at the location of the last 
data point always provides a well constrained rotation 
speed, with no extrapolation of the model fit, 
and it maximizes use of the data on each individual galaxy 
rotation curve.  However, $\Vend$ is difficult to model 
theoretically, because the spatial extent of the
\halpha\ data varies from galaxy to galaxy.  
A common choice for optical rotation curves is the 
rotation speed $V_{2.2}$ at 2.2 disk scale lengths, where the 
rotation curve of a self-gravitating exponential disk would 
peak.  We used this measure in our earlier paper for the 
disk-dominated subset of the galaxy sample (P05).  However, 
for galaxies with significant bulges, 
the value of $R_d$ is sensitive to the degeneracies of 
bulge-disk decomposition, so we do not want to adopt $V_{2.2}$ 
as our primary velocity measure for the full sample.    

After considering a variety of options, we chose 
to evaluate the arc-tangent function at a 
radius ($R_{80}$) containing 80\% of the $i$-band flux. This 
velocity measure, which we refer to as $\Veight$, allows 
a relatively straightforward comparison to 
galaxy formation theories.  For a pure exponential disk,  
$R_{80}$ is 3.03$R_d$, but $R_{80}/R_d$ is smaller for 
galaxies with significant bulges.  The empirical logic of choosing 
$R_{80}$ is evident from Figure~\ref{fig:6090rot}:  most of our 
rotation curves extend close to or beyond $R_{80}$, so 
substantial extrapolation of the rotation curve is rarely 
required, and it is far enough out to be close to $\Vcirc$\ in 
most cases.  

\begin{figure}
\epsscale{0.7}
\plotone{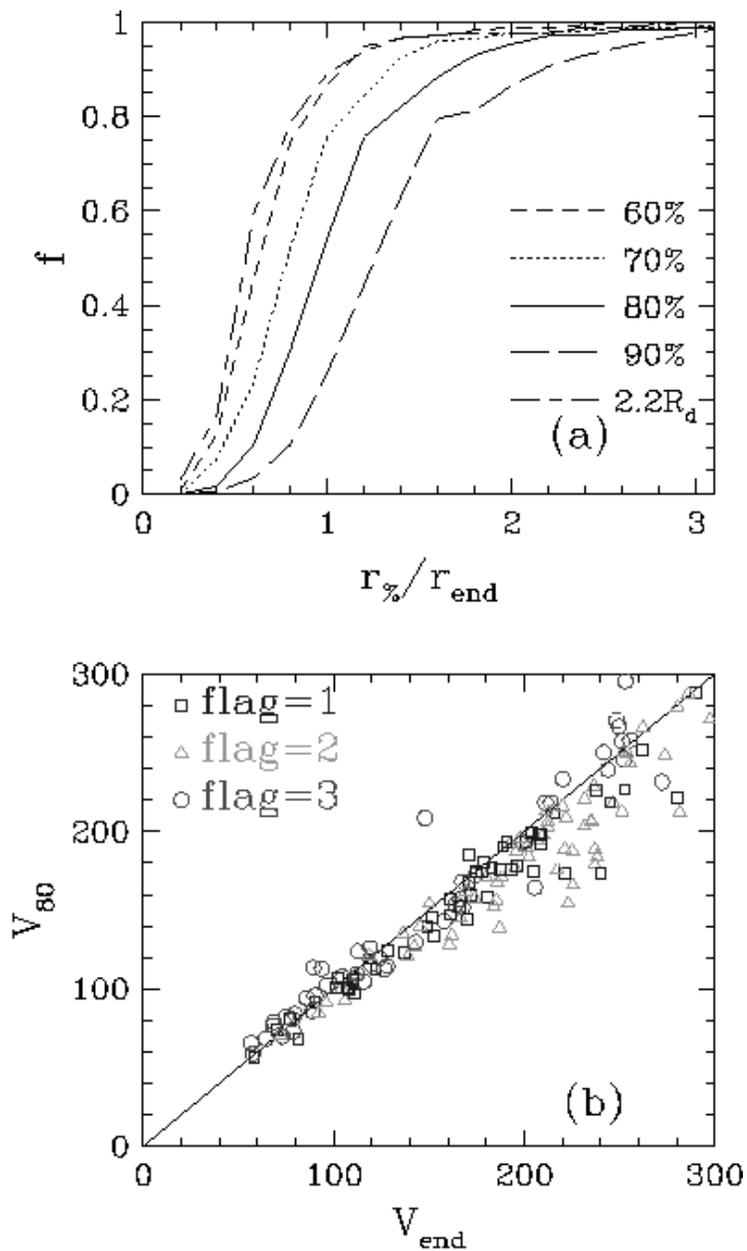}
\caption{(Top) Cumulative distributions of the 
radii that contain 60\%, 70\%, 80\%, and 90\% 
of the $i$-band flux divided by the radius of the last
\halpha\ data point.  The distribution for 2.2$R_d$ 
is also shown.  (Bottom)  Comparison of 
our standard velocity measure $\Veight$, the value
of the arc-tangent function evaluated at the
radius containing 80\% of the $i$-band flux, to $\Vend$, the 
velocity evaluated at the radius of the last \halpha\ 
data point.}
\label{fig:80Vendcum}
\end{figure}

Figure~\ref{fig:80Vendcum}a makes this point quantitatively, showing the cumulative 
distribution of 2.2$R_d$ and the radii containing 60\%, 70\%, 
80\%, and 90\% of the $i$-band flux divided by the radius $\Rend$ 
of the outermost \halpha\ data point.  Half of the galaxies have 
data extending beyond $R_{80}$.  The radii 2.2$R_d$, $R_{60}$, 
and $R_{70}$ would under-utilize the data for a large fraction 
of the sample and be further from $\Vcirc$, while evaluating the 
velocity at $R_{90}$ would require extrapolation in most cases.  
Figure~\ref{fig:80Vendcum}b shows a generally good correlation between $\Veight$\ 
and $\Vend$, with a handful of \flagc\ galaxies having $\Veight$\ 
significantly greater than $\Vend$, and some \flagab\ galaxies 
with extended rotation curves having $\Vend$~$>$~$\Veight$.  Our
choice of $\Veight$\ as a rotation speed definition is very 
similar to that of \cite{per96} and \cite{cat06}, who evaluate the 
rotation curve at the radius 
containing 83\% of total flux (see \cite{cat05} for 
further discussion).  In Appendix A, we present TF 
fits for the alternative velocity definitions $V_{2.2}$ and $\Vend$.

\subsection{Bulge-Disk Decomposition}
Bulge and disk parameters of our target galaxies are needed 
for three reasons: to correct the 
measured velocities ($\Veight$) for disk inclination, to correct
absolute magnitudes for internal extinction by dust in the disk, and
to obtain structural quantities that can be tested for 
correlations with TF residuals.  
The DR2 isophotal axis ratio may be affected by the presence of a 
bulge and spiral arms.  Therefore, refitting the disk, separate from 
the bulge, provides more accurate
 disk axis ratios for the inclination correction of the velocities  
and the internal disk extinction correction.  The bulge-disk decomposition 
also allows measurements of the disk size and 
surface brightness, which are possible third parameters in the TF relation.

We fit a Sersi\'{c} profile \citep{ser68} to the bulge and an  
exponential profile to the disk, using the 
two-dimensional profile fitting program GALFIT \citep{pen02}.  
We run GALFIT on the $g$, $r$, and $i$-band images of 
our galaxies, holding the disk profile fixed as an exponential 
disk and leaving the bulge Sersi\'{c} index $n$ as a free parameter.  
The Sersi\'{c} index describes the central concentration 
of the light profile, with $n$=1 for an exponential disk 
and $n$=4 for a DeVaucouleurs profile. We run GALFIT setting the initial bulge 
Sersi\'{c} concentration equal to $n$=2 and visually inspect 
the results.  The $\chi^2$ is sensitive 
to small-scale asymmetries and variations in the 
galaxy profile that are not two-dimensionally symmetric 
(i.e. HII regions, spiral arms, and bars), so 
visual inspection of the fit results is required to 
ensure that GALFIT does not drift to a local 
$\chi^2$ minimum in an un-realistic 
location of parameter space.  Masking out strongly asymmetric 
features and re-running GALFIT resolves these problems when 
they arise.   

\begin{figure}
\plotone{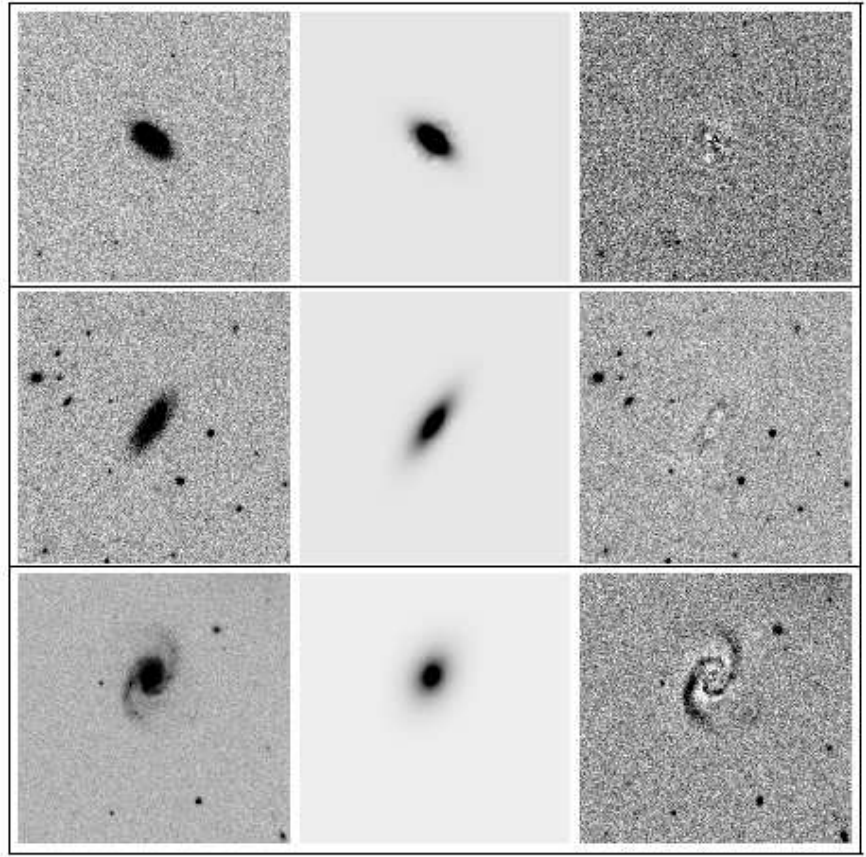}
\caption{Two-dimensional bulge-disk decomposition 
of the $i$-band images of J235656.66-005912.3, 
J094949.62+010533.2, J135946.01+010432.8  (top to 
bottom).  From left to right the panels 
are the data, the GALFIT model, and the residual image given by 
subtracting the GALFIT model from the data.}
\label{fig:ibandgalfit}
\end{figure}

Figure~\ref{fig:ibandgalfit} shows examples of 2-D  
symmetrical Sersi\'{c} profile fits to our \flagabc\ galaxies.  The top and 
middle rows show typical results for galaxies that do not have 
prominent spiral arms.  The bottom panel shows a result for a 
galaxy that has prominent spiral arms.  Since our galaxies are selected 
to be inclined, we rarely find prominent bars in the images.

\begin{figure}
\plotone{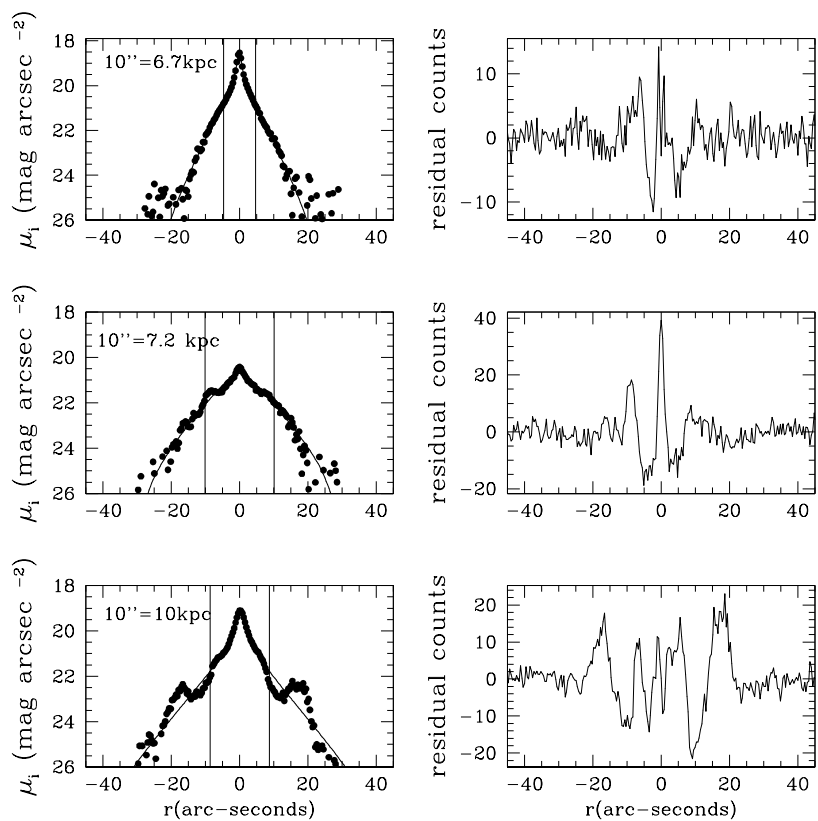}
\caption{Major axis surface brightness profiles 
(averaged over five pixels in each direction) 
of the three galaxies shown in Figure~\ref{fig:ibandgalfit}.  In the left 
panels, points show the data, solid curves show the 
GALFIT model, and vertical lines mark the 
half-light diameter.  Right hand panels 
plot the residual counts (model$-$data); for 
these galaxies, 10 counts/pixel corresponds to roughly 
23.4 mag/arcsec$^2$ in the $i$-band.  }
\label{fig:ibandgalfit1D}
\end{figure}

\begin{figure}
\plotone{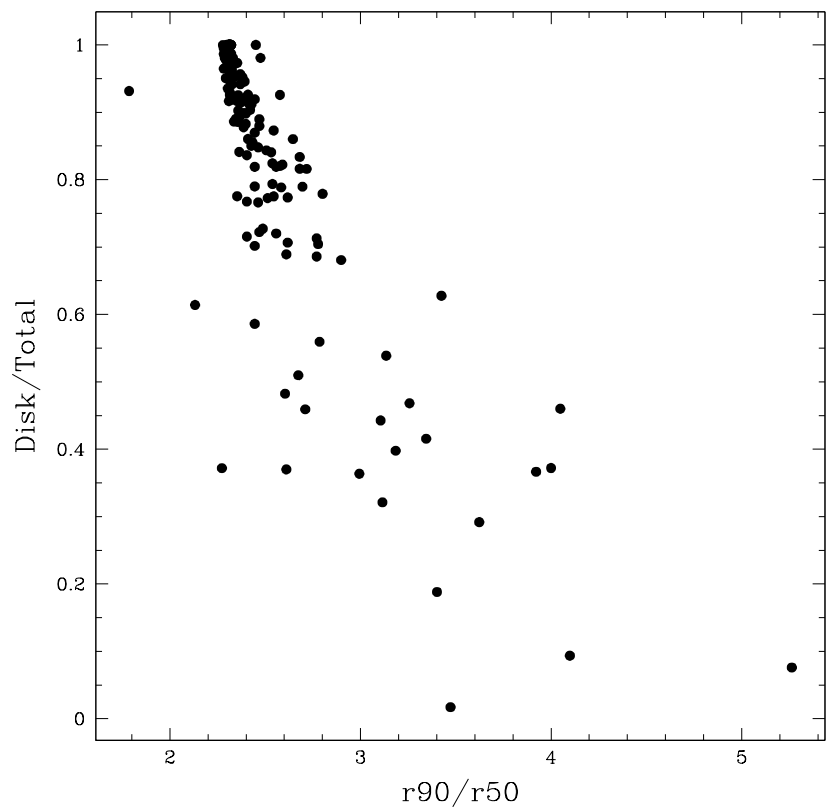}
\caption{The $i$-band $D/T$ vs. the concentration index ($r_{90} / r_{50}$) for the 
$\flagabc$ galaxies.
  }
\label{fig:dtihub}
\end{figure}

Figure~\ref{fig:ibandgalfit1D} shows 1-D profiles along the major 
axis of the data, model, and residual images for galaxies shown in Figure~\ref{fig:ibandgalfit}.  
From Figure~\ref{fig:ibandgalfit1D} we can conclude that our best-fit 
models accurately describe the galaxy surface 
brightness profiles down to $\sim$25 mag/arcsec$^2$ or 
fainter, except for small scale variations.  The bulge-to-disk 
flux ratio, disk axis ratio, and disk position angle
 are generally robust outcomes of the fitting procedure, 
converging quickly with little sensitivity 
to the initial parameter choice. The bulge radius, 
Sersi\'{c} index, and position 
angle are subject to more uncertainty, in part 
because the bulge component is often 
similar in size to the seeing FWHM.  Figure~\ref{fig:dtihub} shows that
approximately 20\% of our galaxies with usable rotation curves have significant bulges, 
where early-type galaxies have concentrations around 2.6.  Figure~\ref{fig:dtihub} 
also shows that galaxies with lower concentrations tend to 
have higher disk-to-total flux fractions.  A visual inspection 
of the \flagabcd\ galaxies confirms this trend, where 
the \flagcd\ galaxies tend to have early morphological 
types.  The \flaga\ galaxies are predominantly Sb and Sc 
Hubble types, whereas the \flagd\ galaxies are predominantly 
Sa with a few Sc galaxies.  The \flagbc\ galaxies are mixed 
between Sa, Sb, and Sc types.

\subsection{Internal Extinction Corrections}
Dust in galaxy disks absorbs a larger fraction of the
disk light in edge-on directions.   Therefore, 
it is standard practice to apply internal extinction corrections to 
luminosities used for the TF relation.  We 
follow this practice here and use the \cite{tul98} 
formulation adapted to our bands.  We apply 
internal extinction corrections to 
the disk, while assuming that the bulge 
has no correction to the face-on value because it has little dust. 
\cite{tul98} provide prescriptions for the internal extinction 
corrections in the Johnson $B$ (438 nm), $R$ (641nm), $I$ (798 nm), 
and $K'$ filters as a function of the galaxy inclination 
and the absolute magnitude in that band.  We convert 
the SDSS $g$ (469 nm), $r$ (617nm), $i$ (748nm), and $z$-band (893 nm)
absolute magnitudes to Johnson magnitudes using 
the conversions in Table 7 of \cite{smi02}, then 
linearly interpolate the \cite{tul98} corrections back 
from the Johnson central wavelengths to the SDSS 
central wavelengths as listed above.  This internal 
extinction correction is then applied to the disk 
$g$, $r$, $i$, and $z$-band disk fluxes, which are then
 added to the un-corrected bulge fluxes.

\begin{figure}
\plotone{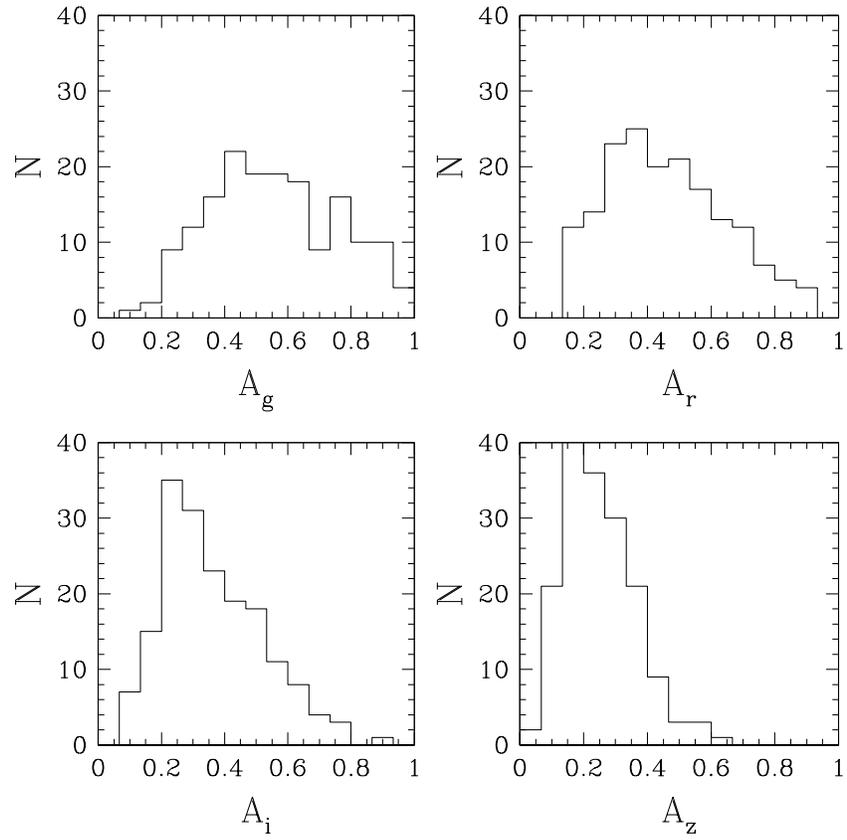}
\caption{Histograms of the total internal extinction for 
the $g$, $r$, $i$, and $z$-bands. 
\label{fig:ainttotal}}
\end{figure}

\begin{figure}
\epsscale{1.0}
\plotone{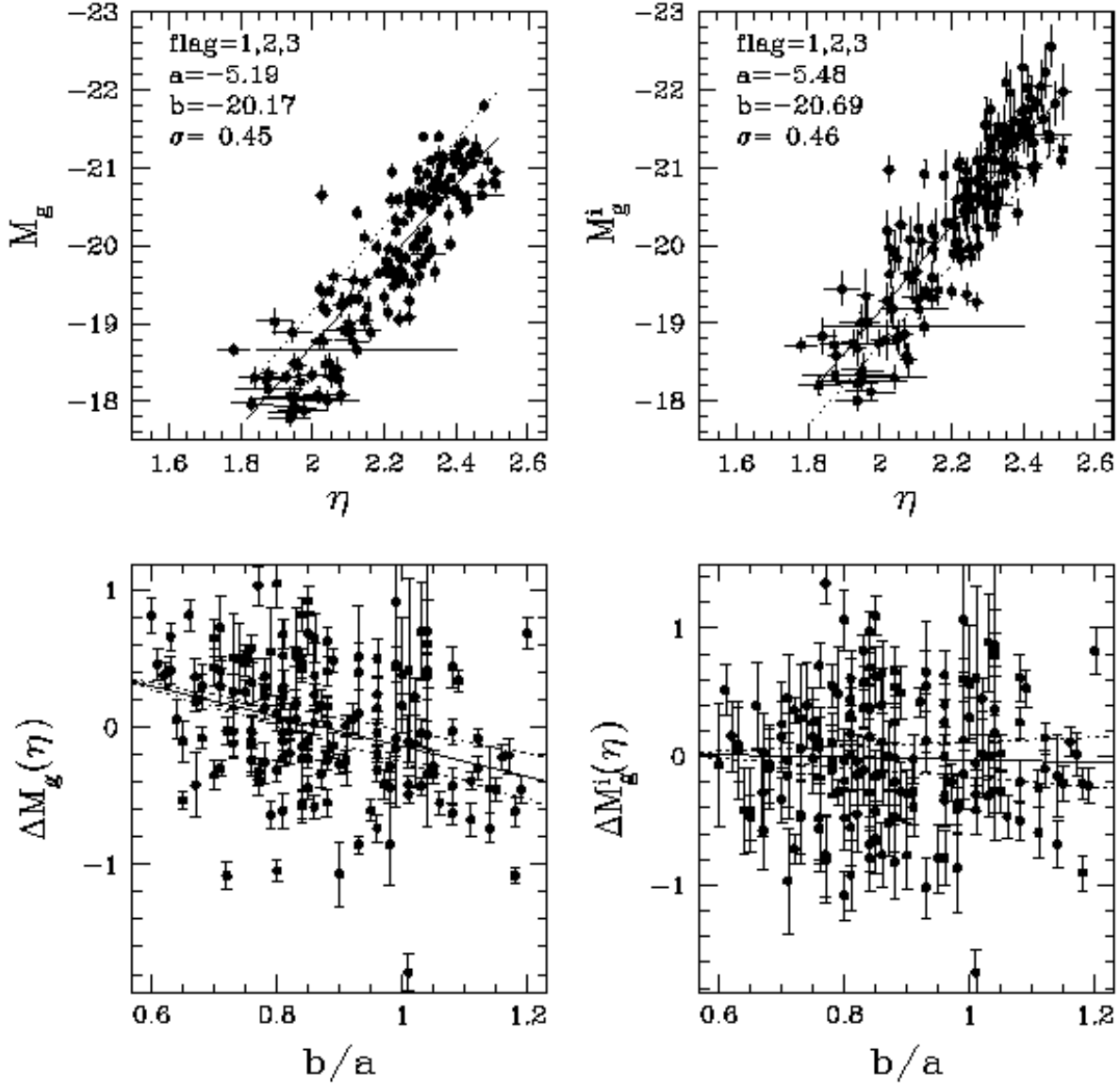}
\caption{Effect of the internal extinction correction 
on the $g$-band TF relation.  Upper panels show the data (points) 
and best-fit TF relations (solid lines) without (left) and
with (right) internal extinction corrections.  The dotted line 
shows the fit from the other panel.  Lower panels plot TF 
residual against $i$-band axis ratio, with solid lines showing 
the best-fit trend and dotted lines the bootstrap-estimated 
uncertainty. 
We define 
$\eta \equiv \log \Veight/(\kms)$, where $\Veight$ is the
inclination corrected circular velocity at the radius 
containing 80\% of the $i$-band flux.}
\label{fig:gcompare}
\end{figure}

Figure~\ref{fig:ainttotal} shows histograms of the total 
internal extinction correction in each band.  
Typical values of the $g$-band internal extinction 
range from 0.2 to 0.8 magnitudes.  The median internal 
extinction correction decreases from 0.6 in the $g$-band
to 0.2 in the $z$-band.  Figure~\ref{fig:gcompare} shows 
the $g$-band TF relation with and without the 
internal extinction correction.  Solid lines show fits 
using the maximum likelihood procedure discussed in 
\S 5 below.  The slope of the extinction-corrected TF relation is similar 
to the uncorrected slope ($-5.5 \pm 0.2$ versus $-5.2\pm 0.2$), 
though of course the extinction correction increases  
the zero point by the average
value of the extinction.  The lower panels of Figure~\ref{fig:gcompare}
show that the internal extinction correction successfully 
removes a weak trend of TF residual with axis 
ratio, so it appears to be a valid and useful correction 
on average, even if it is uncertain on a galaxy by galaxy 
basis.  We assume, somewhat arbitrarily, that the 
inclination correction {\it uncertainty} is 1/3 of 
the correction itself.  Results in Appendix A show
 that, with this assumed observational 
uncertainty, the internal extinction 
corrections do not change the inferred intrinsic scatter $\sigma$.  
All magnitudes discussed in the remainder of the paper have this 
correction applied. 

\subsection{Error Budget}
We are ultimately interested in the slope, 
intercept, and intrinsic scatter of the TF 
relation.  To estimate these, we need an accurate 
characterization of the observational 
errors of our data.  This is especially 
important for estimating intrinsic scatter, since 
the total scatter about the fitted relation is, 
roughly speaking, the quadrature sum of the intrinsic 
scatter and the observational errors.  

The TF relation is the correlation of the logarithm of the rotation velocity with
the absolute magnitude.  We define the logarithm of 
the rotation velocity to be
\begin{equation}
\eta \equiv \log \left (\frac{\Veight}{\sin i  (1+z)} \right),
\end{equation}
where $i$ is the disk inclination angle inferred from the bulge-disk
decomposition procedure described in \S 4.3, and
the $(1+z)$ term accounts for cosmological broadening.  We 
relate the inclination to the observed axis ratio using the equation
\begin{equation}
\sin i = \sqrt{ \frac{1-(b/a)^2}{1-0.19^2}},
\end{equation}
where $b/a$ is the $i$-band disk axis ratio 
determined using GALFIT, and $0.19$ takes into account 
the finite thickness of the disk (see the discussion 
in \cite{hay84}); varying $0.19$ over the range 0.10$-$0.25 causes the 
inclination correction to vary by $<$ 1.5\%.  We define the absolute magnitude to be 
\begin{equation}
M_\lambda \equiv m_{{\rm Petro},\lambda} - 
 5\log_{10}(d_{\rm lg}/(1+z)) + 
 5 - A_{\lambda}^i - A_{\lambda}^{\rm MW} + K_{\rm corr,\lambda},
\end{equation}
where $\lambda$ denotes the band ($g$, $r$, $i$, or $z$), 
$m_{{\rm Petro},\lambda}$ is the SDSS Petrosian magnitude in 
the $\lambda$-band,
$A_{\lambda}^i$ is the internal extinction correction described
in \S 4.4, $A_{\lambda}^{\rm MW}$ is the 
correction for Milky Way extinction taken from the SDSS database
(based on \citealt{sch98}), $K_{\rm corr,\lambda}$ is the 
K-correction from \cite{bla03b}, and $d_{\rm lg}$ is the 
luminosity distance to the galaxy from the Local Group.  As noted 
earlier, we compute distances using the 
SDSS heliocentric redshifts corrected to the rest frame 
of the Local Group barycenter (\citealt{yah77}),
assuming a cosmological model with $\Omega_m=0.3$, 
$\Omega_{\lambda}=0.7$, and $h = 0.7$.  There are six
galaxies (J233152.99-004934.4, 
J144418.37+000238.5, J095555.07-001125.0, J112346.06-010559.4,
J203523.80-061437.9, J001006.62-002609.6) that  have poorly 
estimated Petrosian magnitudes 
in the DR2 photometric pipeline.  For these galaxies, we use
the total Sersi\'{c} magnitude measured using GALFIT.

The uncertainty in the absolute magnitude is calculated 
using standard propagation of errors, 
\begin{equation}
(\delta M_{\lambda})^2 = (\delta m_{\rm petro,\lambda})^2 + \left (5\times0.434\frac{\delta z}{z} \right)^2 +
\left (5\times0.434\frac{\delta V_{\rm pec}}{cz} \right)^2 + (\delta A_{\lambda}^i)^2 ,
\end{equation}
where 0.434 converts from natural to base-10 logarithms.  
Uncertainties in apparent magnitude $\delta m_{\rm petro,\lambda}$ 
and redshift $\delta z$ are taken from the DR2 database, though 
we impose a minimum redshift error of $c\delta z$ $=$ 30 $\kms$.  
The uncertainty in the Milky Way foreground extinction, 
$A_{\lambda}^{MW}$, is assumed to be negligible.  We incorporate a distance 
uncertainty corresponding to $\delta V_{\rm pec}$ = $300$ $\kms$ 
when calculating disk scale 
length and luminosity uncertainties, to account 
for the typical amplitude of small scale peculiar 
velocities (\citealt{str95}).  We assume the uncertainty 
due to the internal extinction correction is one third 
of the calculated value, $\delta A_{\lambda}^i$ = $A_{\lambda}^i$/3, and we 
ignore any (much smaller) uncertainty due to the 
uncertainty in the GALFIT determined $i$-band axis 
ratio.  In practice, peculiar velocity and 
internal extinction uncertainties completely dominate the uncertainty 
in $M_{\lambda}$.  The 10-th, median, and 90-th percentile 
values of the peculiar velocity uncertainty, 
$5\times0.434\frac{\delta V_{\rm pec}}{cz}$, are 0.049, 0.080, 
and 0.118, respectively.  The 10-th, median, and 90-th percentile 
total $i$-band internal extinction correction uncertainties 
are 0.041 mag, 0.088 mag, and 0.16 mag, respectively.    

The velocity width has uncertainties due to 
the inclination correction, measurement of 
$\Veight$ from the rotation curve, and possible 
systematic uncertainties due to slit misalignment.  
As described in \S 2, we removed six galaxies from 
our sample for which misalignment 
could cause a $>$10\% change in the circular velocity; we
do not correct six other galaxies with slight slit misalignments 
for which the correction is $<$10\%.  The uncertainty in the 
inclination corrected velocity width ($V_{80,i}$) 
due to the inclination uncertainty 
and uncertainties in the measurement of $\Veight$ is
\begin{equation}
\left (\frac{\delta V_{80,i}}{V_{80,i}}\right)^2 = \left (\frac{\delta \Veight}{\Veight} \right)^2 + 
\left (\frac{\delta(b/a) (b/a)}{1-(b/a)^2} \right )^2 ~.
\end{equation}
 The first term dominates for essentially all of the galaxies.  
 Using the covariance matrix returned by the
 Levenberg-Marquardt method, the uncertainty of the arc-tangent 
function measured at $R_{80}$ is related to the uncertainty in the arc-tangent 
function parameters via the equation
\begin{equation}
\delta \Veight^2 = \left (\sum_{ij} C_{ij} \frac{\partial^2 V_{\rm arctan}(R_{80}|a_i,a_j)}{\partial a_i \partial a_j} \right)^2.
\end{equation}
The indices run from 1 to 4 for the four arc-tangent 
function parameters $a_i$ and $a_j$.  The axis ratio 
is determined from the disk component after running 
GALFIT on the $i$-band corrected frames, with the uncertainty 
reported by GALFIT.  
The 10th-percentile, median, and 90th-percentile 1$\sigma$ uncertainties 
in log $V_{80,i}$ are 0.006, 0.011, and 0.044 ${\rm log_{10}}$(km/sec) 
respectively.  For a TF slope of $-6.0$ mag / ${\rm log_{10}}$ (km/sec), 
these uncertainties correspond to 0.034, 0.063, and 0.264 mag
respectively.  A typical uncertainty for our program galaxies is 
thus $\sim$ 0.063 mag from the $V_{80,i}$ uncertainty, $\sim$ 0.088 mag 
from the internal extinction correction uncertainty, and $\sim$ 0.080 mag 
from the peculiar velocity, summing in quadrature to  0.13 mag.
This is much smaller than the intrinsic 
scatter measured in \S 5.2.  However, the 
variation from galaxy to galaxy is large.
For the rest of the paper we will use inclination corrected 
velocities and drop the subscript $i$ from $V_{80,i}$.  Table 2
summarizes the photometric parameters and their errors, and Table 3
summarizes the velocity-width measurements and their errors.

\section{The Tully-Fisher Relation}
\subsection{Modeling the TF Relation}
We use a maximum likelihood method to estimate 
the slope $a$, intercept $b$, and intrinsic 
scatter $\sigma$ of the TF relation.  For the 
``forward'' relation, the independent and 
dependent variables are $x$ = $\eta$ $\equiv$ ${\rm log} (\frac{\Veight}{{\rm sin {\it i} (1+z)}})$  
and $y$ = $M_{\lambda}$, respectively.  We assume that the intrinsic 
scatter is Gaussian in form, so that the probability 
that galaxy $i$ has a {\it true} absolute magnitude 
$y_i$ given a {\it true} velocity width $x_i$ is
\begin{equation}
p(y_i|x_i) = (2\pi\sigma^2)^{-1/2} {\rm exp} \left [\frac{-(y_i - \overline{y_i})^2}{2\sigma^2}\right],  
\end{equation}
where
\begin{equation}
\overline{y_i}=a(x_i-x_0) + b
\end{equation}
is the mean value expected for a linear TF relation with 
parameters $a$ and $b$.  The value of $x_0$ is chosen 
so that there is no correlation (or very little 
correlation) between the statistical errors in 
$a$ and $b$.  For the ``inverse'' TF relation, we 
adopt the same model but with $x$ = $M_{\lambda}$ and $y$ = $\eta$.  
The forward and inverse relations correspond to different 
assumptions about which physical parameter is ``primary'':  
in the forward relation, the absolute magnitude has Gaussian 
scatter about a mean value determined by the linewidth, and the 
reverse holds for the inverse relation.  

We assume that the observational estimates of 
$\hat{x_i}$ and $\hat{y_i}$ are Gaussian distributed 
about the true values of $x_i$, $y_i$, i.e., 
\begin{equation}
p(\hat{y_i}|y_i) = (2\pi\sigma_{y,i}^2)^{-1/2} {\rm exp} \left [\frac{-(\hat{y_i}-y_i)^2}{2\sigma_{y,i}^2}\right]
\end{equation}
and 
\begin{equation}
p(\hat{x_i}|x_i) = (2\pi\sigma_{x,i}^2)^{-1/2} {\rm exp} \left [\frac{-(\hat{x_i}-x_i)^2}{2\sigma_{x,i}^2}\right],
\end{equation}
where $\sigma_{x,i}$, and $\sigma_{y,i}$ are the 
measurement uncertainties determined as described in \S 4.5.  
The values of $a$, $b$, and $\sigma$ can then be estimated by 
maximizing the log-likelihood
\begin{equation}
{\rm ln}\,L = \sum_i {\rm ln}\, p(\hat{y_i}|\hat{x_i},\sigma_{x,i},\sigma_{y,i}). 
\end{equation}
The likelihood for an individual data point can be written
\begin{eqnarray}
p(\hat{y_i}|\hat{x_i})  & = &
\int_{-\infty}^{\infty} dy_i \, p(\hat{y_i}|y_i)p(y_i|\hat{x_i}) , \nonumber \\
  & = & \int_{-\infty}^{\infty} dy_i \, p(\hat{y_i}|y_i) \int_{-\infty}^{\infty} dx_i \, p(y_i|x_i) p(x_i|\hat{x_i}) .
\end{eqnarray}
We set $p(x_i|\hat{x_i})$ $=$ $p(\hat{x_i}|x_i)$, implicitly 
assuming a flat prior for $p(x_i)$ over the (typically narrow) 
range of values allowed by the uncertainty in $\hat{x_i}$.  
Substituting equations (8)-(11) and (13) into (12) and 
simplifying yields the expression
\begin{equation}
{\rm ln}\, L = -\frac{1}{2} 
  \sum_i {\rm ln}\, (\sigma^2 + \sigma_{y,i}^2 + a^2\sigma_{x,i}^2) - 
  \sum_i \frac{[\hat{y_i}-(a\hat{x_i}+b)]^2}{2(\sigma^2+\sigma_{y,i}^2+a^2\sigma_{x,i}^2)}
  +\,{\rm constant}.
\label{eqn:like}
\end{equation} 
Most previous TF studies have implicitly set $\sigma$ to zero when 
finding the best-fit values of $a$ and $b$, then estimated 
$\sigma$ after the fact from the difference between the observed 
scatter and the estimated contribution of observational errors.  In 
the presence of non-zero intrinsic scatter, this method 
assigns too much weight to the data points with the 
smallest observational errors, yielding statically non-optimal estimates of $a$ 
and $b$ and underestimates of their uncertainties.  For example, two 
data points with near-zero errors would completely dominate the fit, 
while in fact they should be weighted by 1/$\sigma^2$.  Further 
discussion of these issues can be found in \cite{dag05}.

The equation $\delta {\rm ln} L / \delta {\rm ln} b$ $=$ 0 (see 
eq. 14) can be solved 
analytically given values of $a$, $\sigma$, and the input data 
($\hat{x_i}, \hat{y_i}, \sigma_{x,i}, \sigma_{y,i}$).  
We determine the maximum likelihood parameters by performing a grid search 
in $a$ and $\sigma$, finding the best-fit value of 
$b$ (analytically) for each ($a$, $\sigma$) combination, 
then choosing the ($a$, $b$, $\sigma$) combination that maximizes 
ln$L$.  We determine the 1$\sigma$ errors on $a$, $b$, and $\sigma$ 
by repeating this procedure for 100 bootstrap subsamples of the full 
data set, taking the dispersion among the bootstrap estimates as 
the uncertainty in the parameter.  We choose the values of $x_0$ in equation 
(9) so that there is essentially no covariance between the error 
in $a$ and the error in $b$; specifically we choose the value of $x_0$ 
so that fixing $a$ to a value of $\pm$1$\sigma$ from its best-fit 
value does not change the best-fit value of $b$.  The resulting zero-points 
in the $g$, $r$, $i$, and $z$ TF fits are, respectively, $\eta_0$ 
= 2.22, 2.22, 2.22, 2.23 for the forward fit and 
$M_{\lambda_0}$ = $-20.607$, $-21.107$, $-21.327$, and $-21.400$ 
for the inverse fit.  

\subsection{The TF Relation in the SDSS Bands}
\begin{figure}
\epsscale{1.1}
\plotone{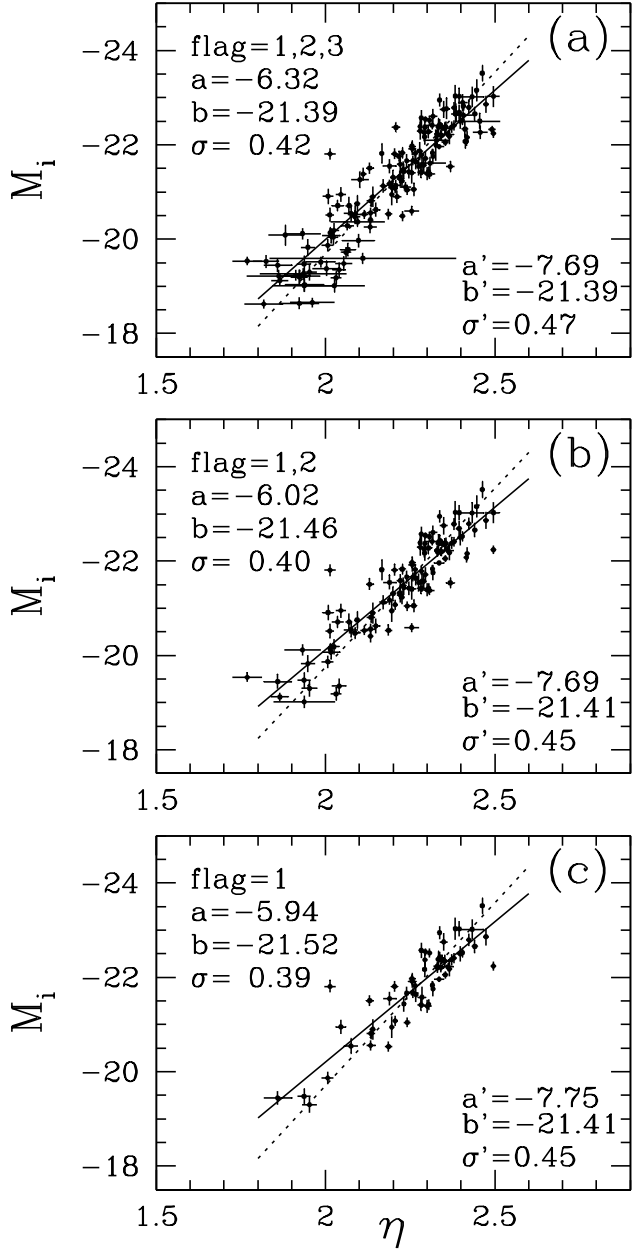}
\caption{
The $i$-band TF relation for the
full galaxy sample (top), \flagab\ 
galaxies (middle), and \flaga\ galaxies 
only (bottom).  Solid and dotted lines show 
forward and inverse fits, respectively.  The forward 
slope, intercept (at $\eta_0$ $=$ 2.22), and 
intrinsic scatter are listed as $a$, $b$, and $\sigma$, 
while $a'$, $b'$, $\sigma'$ refer to the converted 
inverse fit parameters.  Parameter uncertainties 
are listed in Tables~\ref{tbl:tf} and~\ref{tbl:appx}.
Table~\ref{tbl:tf}.} 
\label{fig:ibandTF}
\end{figure}

Figure~\ref{fig:ibandTF}a shows the $i$-band TF relation with 162 \flagabc\ 
data points.  The data show a clear linear trend of $\Mi$ 
with $\eta$ and an 
intrinsic scatter larger than the error bars.  The best-fit 
parameters for the forward relation are $a = -6.32 \pm 0.22$,
 $b = -21.390 \pm 0.035$ (at $\Veight$ = 166.0 $\kms$), and 
intrinsic scatter $\sigma = 0.423 \pm 0.035$.\footnote{The units 
are mag/log$_{10}$ ($\kms$) for the slope and mag for the 
intercept and scatter.  For brevity, we will omit these units 
in the text.}  Table~\ref{tbl:tf}
lists these parameters and the parameters $a_{\rm inv}$, 
$b_{\rm inv}$, $\sigma_{\rm inv}$ of the inverse fit.  
When comparing the inverse TF to the forward 
TF relation in the figures and text, 
we invert the inverse relation and refer to 
the slope and scatter by $a'$ = $1/a_{\rm inv}$ and 
$\sigma'$ = $\sigma_{\rm inv}$/$a_{\rm inv}$, and quote the intercept
$b'$ as the value of $M_{\lambda}$ that corresponds to 
the zero-point $\eta_0$ used in the forward fit.  The 
quantities $a'$, $b'$, $\sigma'$ have the same
units as $a$, $b$, $\sigma$ and can be directly compared.  For 
the \flagabc\ sample, the inverse TF $i$-band relation $a'$, $b'$, 
and $\sigma'$ are $-7.69 \pm 0.29$, $-21.39 \pm 0.04$, 
and $0.47 \pm 0.04$.  For all bands and samples, the inverse 
relation has a steeper slope than the forward 
relation but similar scatter (see discussion in 
\S 5.3 below).  Throughout the paper we use the 
terms ``steep'' and ``shallow'' in reference to the 
slope (more negative is steeper) of the forward TF 
relation ($\Veight$ predicting $M_{\lambda}$).

Figure~\ref{fig:ibandTF}b,c shows the $i$-band TF relation when less spatially 
extended rotation curves (\flagbc) are removed.  As \flagc\ and then 
\flagb\ galaxies are removed, the slope becomes slightly shallower, 
changing from $-6.32 \pm 0.22$ to $-6.02 \pm 0.32$ to $-5.94 \pm 0.44$.  
The change is at the $\sim$ 1$\sigma$ level, and it is driven largely 
by the change in the sample luminosity distribution, as there are 
fewer \flagab\ galaxies at low luminosity.  Similar 
trends are seen in $g$, $r$, and $z$.  The intrinsic scatter 
declines by a statistically insignificant amount, from 0.42$\pm$0.04 
to 0.39$\pm$0.05 from \flagabc\ to \flaga\ only.  The constancy 
of $\sigma$ implies that our fitting procedure gives a reasonable 
estimate of the uncertainty in $\Veight$\ even for rotation 
curves that are rising at the outermost point.  Given the 
insensitivity of the results to 
including \flagbc\ galaxies, we will use the full sample 
henceforth, for improved statistics and greater sample completeness.  
We list TF parameters for the \flagab\ sample in Table~\ref{tbl:appx} 
(see Appendix A).    

Figure~\ref{fig:griz} shows the TF relation for the $g$, $r$, $i$ and $z$-bands, with 
parameters fit to the data as outlined in \S 5.1.
The forward TF slope increases towards the redder bands, from $-5.48$ to 
$-6.59$, a change much larger than the statistical slope uncertainties 
(typically 0.2).  This is the expected trend, as fainter galaxies 
typically have bluer colors.  Trends in the intercept are not particularly 
meaningful, as they depend on the adopted (AB) magnitude system and 
stellar mass-to-light ratios in different wavebands.  The inverse 
TF slopes are always steeper than the forward slopes, and they
 show the same trend with wavelength.  
The intrinsic scatter is slightly larger in the $g$-band, and roughly 
constant in $r$, $i$, and $z$.  This constancy implies that the extinction 
corrections are not an important source of 
scatter.  It further suggests that the 
intrinsic scatter is not dominated by variations 
in stellar populations, a point we will 
return to in \S 6 below.  Excluding \flagc\ and 
\flagb\ galaxies in the $g$, $r$, and $z$-bands shows the same trends 
seen in Figure~\ref{fig:ibandTF}.  TF fits using velocity width 
definitions $V_{2.2}$ and $\Vend$ are discussed in Appendix A.  

As shown in \S 2 (cf. Figure~\ref{fig:histogram}), the $\Mr$ distribution 
of our sample is approximately flat in the range $-18.5 > \Mr > -22$, 
but not exactly so.  In order to provide a well defined target 
for galaxy formation theories, we have also computed TF parameters 
after weighting galaxies to obtain the results expected for a truly 
flat $\Mr$ distribution.  Using the plotted distribution of galaxies 
in 0.5-magnitude bins, we give each galaxy a weight $w_i$ defined as 
the ratio of the number of galaxies in the most populated bin to 
the number of galaxies in its bin.  These weights are incorporated 
into the fitting procedure by multiplying each term inside 
the sums of equation (14) by $w_i$, as though we had observed
the galaxy $w_i$ times ($w_i$ goes inside the ln of the 
first term).  The weights vary from 1 to 5, 
depending on the bin.  We do not include the few galaxies 
with $\Mr$ $>$ $-18.5$ and $\Mr$ $<$ $-22$.  The results are 
quoted in Table~\ref{tbl:appx}.
Overall, the parameters change very little
when weighted by our observed $M_r$ distribution.  
We have carried out a similar weighting experiment to examine 
the effect of exactly reproducing the $g$-$r$ color distribution 
of the DR4 ``control'' sample shown by the dotted curve 
in Figure~\ref{fig:cumdist}a.  
The change in the TF parameters (not listed) is much smaller than 
our statistical errors.

Our sample has absolute magnitude cuts that 
can create Malmquist-type biases in TF estimates.  The explicit 
absolute magnitude-redshift cuts outlined in \S 2, and the 
\flagabc\ absolute magnitude distribution seen in Figure~\ref{fig:histogram}, 
may cause a shallowing of a TF slope and reduction of the intrinsic
scatter when these cuts are applied to data having non-zero scatter.  
Random errors in galaxy distances and scatter across selection 
boundaries, due to peculiar velocities typically of the 
order 300 $\kms$, can also induce Malmquist-type biases in TF estimates.
Appendix B describes a Monte Carlo experiment aimed at testing for 
such biases inherent in our sample, assuming a power-law TF 
relation with Gaussian intrinsic scatter.  
In short, generating a Monte Carlo sample 
with an input $a$ $=$ -7.35, $b$ $=$ -21.27, and $\sigma$ $=$ 0.58, then 
applying the selection criteria outlined above, yields 
a best-fit forward relation of $a$ $=$ -6.32, $b$ $=$ -21.39, and $\sigma$ $=$ 0.42 
which is close to the observed $i$-band TF relation (Table 4).  The 
difference between the input and best-fit forward relations is 
several times our statistical uncertainties.  The 
difference between the inverse relations is small.
In this paper we present a TF relation for a sample of galaxies with 
well defined selection criteria, enabling a theory to reproduce our 
sample, using our selection criteria, in an attempt to model our measured
TF relation.  Table 4 represents the TF parameters measured with
our selection criteria.  

\subsection{Intrinsic Scatter}
One of our principal objectives is a robust estimate 
of the intrinsic scatter of the TF relation for a broadly selected sample 
of galaxies.  Previous observational studies have shown a very 
wide range of estimates, with some as low as 0.1 to 0.15 mags \citep{ber94}.  
For a large sample of late-type spirals, with 
similar analysis methods to those used here, C97  
reports 0.46 mag of total (intrinsic $+$ observational) 
 $r$-band scatter.  
K02 calculates and $R$-band intrinsic scatter of 
0.4 mag for the broadly selected Nearby 
Field Galaxy Survey \citep{jan01} and Ursa Major 
samples, after the selection criteria are matched.
 However, V01 reports a much smaller scatter, 
0.15 $-$ 0.18 mag in $R$-band, for galaxies in the 
Ursa Major cluster, using the flat part of the HI 
rotation curve as the circular 
velocity measure (see Table 6 of V01).  This 
estimate arises after subtracting 0.17 mag of scatter
attributed to the depth of the cluster, and 
changes to the sample or velocity width definition can boost the 
$R$-band scatter in the V01 analysis as high as 0.54 mag.  Observational 
estimates from other studies span the full range of 
values quoted above (see e.g., the review by \citealt{str95}).

Many papers do not clearly distinguish the intrinsic scatter from 
the total scatter.  In part, the observational errors themselves 
can be difficult to estimate for nearby samples in which distance 
errors dominate and are difficult to define precisely.  
In addition, previous studies have generally not estimated the 
intrinsic scatter as part of the TF fitting procedure as we have 
done, but have rather subtracted a ``typical'' observational 
error in quadrature from the total scatter.

A crucial feature of our sample is that the typical 
observational errors are smaller than the intrinsic scatter because
we are measuring galaxies with $V$~$>$~$5000$~$\kms$ and have photometry 
that allows precise estimates of magnitudes and axis ratios.
The most debatable elements of our observational error budget (see 
\S 4.5) are the $300 \kms$
peculiar velocity error and the internal extinction error of $1/3$ the 
internal extinction correction.  These typically contribute 0.080 and 
0.088 mag to the absolute magnitude error.  If we increase the 
assumed peculiar velocity uncertainty from 300 to 500 $\kms$, per
galaxy, our estimate of the $i$-band intrinsic scatter decreases 
from 0.42 to 0.40 mag.  If we set the assumed internal 
extinction error to zero, the estimated intrinsic scatter 
increases from 0.42 to 0.43 mag.  Thus, our estimate of 
$\sigma$ is relatively insensitive to the uncertain elements of our 
error budget.  Even if we take the drastic step of setting all 
of our observational errors to zero, the best-fit intrinsic 
scatter only increases to 0.48 mag.     

When computing inclination corrected velocities, we take the 
uncertainty in $\sin i$ from the uncertainty in the 
axis ratio of the GALFIT bulge-disk decomposition.   
We thus implicitly assume that the disk is adequately 
described by a circularly symmetric exponential, and disk 
ellipticities by definition contribute to the intrinsic 
rather than the observational scatter.  \cite{zar95}, 
from a study of face-on galaxies, estimate typical ellipticities 
of $\sim$ 0.05 for the gravitational 
potential in the disk plane, which would cause $\sim$ 0.15 mag of TF 
scatter from a combination of inclination correction errors and 
non-circular motions (see also \citealt{fra92,ryd05}).    

\begin{figure}
\plotone{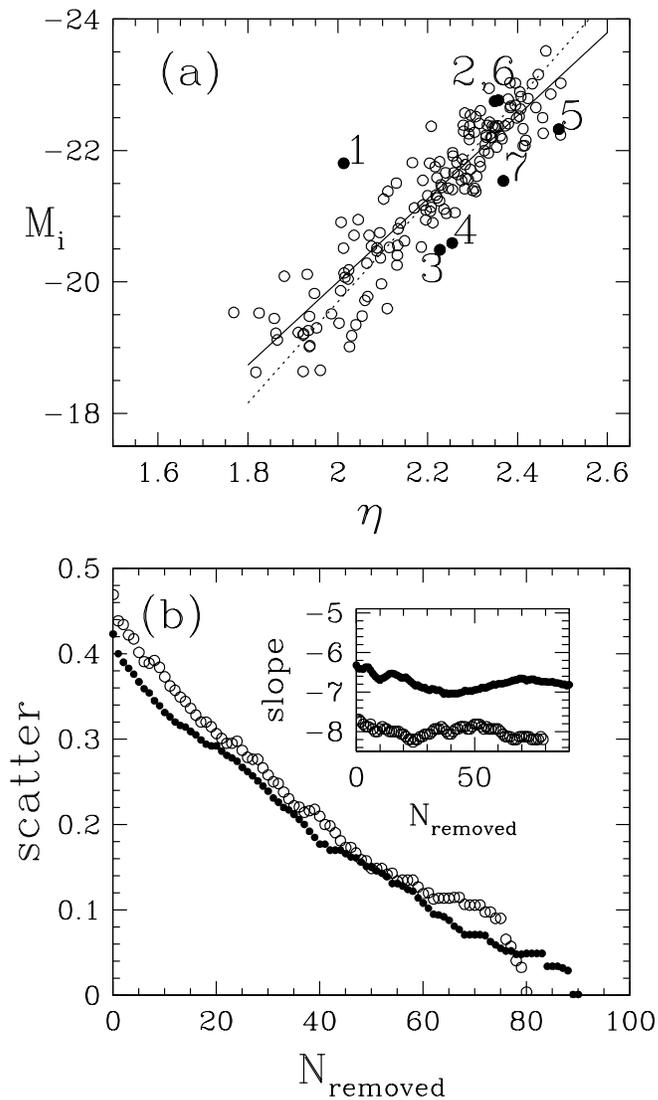}
\caption{
(Top) TF relation with the seven largest outliers 
(in terms of $\Delta\chi^2$) marked in rank order.
(Bottom) Estimated intrinsic scatter of the $i$-band
TF relation after omitting the $N_{\rm removed}$
data points with the largest $\Delta \chi^2$.
The solid circles represent the forward TF relation 
and the open circles the inverse.
The slope as a function of removed 
data points is shown in the inset.}
\label{fig:iband.V80.deltedb}
\end{figure}

Since we have broader sample selection than most previous studies, 
it is interesting to ask whether outliers have an important 
impact on the TF scatter (or other TF parameters).  
Figure~\ref{fig:iband.V80.deltedb}a shows the $i$-band TF 
relation with the seven top contributors 
to $\chi^2$ marked.  From largest to smallest $\Delta$$\chi^2$
 they are J021941.13-001520.4, J235106.25+010324.0, 
J124428.85-002710.5, J204913.40+001931.0, J005650.61+002047.1, 
J013142.14-005559.9, and J235607.82+003258.1.  None of these 
galaxies show obvious rotation curve anomalies, but some are 
morphologically unusual.  J021941.13-001520.4 has extended 
and asymmetrical spiral arms, and its GALFIT axis ratio 
is 0.51, giving it a substantial inclination correction.  It is 
therefore possible that the low rotation speed of this galaxy 
arises from an inaccurate inclination correction.  The second 
galaxy, J235106.25+010324.0, has normal morphology, as does the fifth galaxy 
 J005650.61+002047.1.  The third galaxy, J124428.85-002710.5, has 
a prominent central point source that produces a persistent 
residual in the bulge-disk decomposition.  However, it is under-luminous 
relative to the mean TF relation, so an AGN contribution 
cannot explain the anomaly.  The fourth galaxy, J204913.40+001931.0,  
has a prominent dust lane, as does the sixth, J013142.14-005559.9.  
Finally J235607.82+003258.1 has a large ring $\sim$ 28 kpc 
in size centered on the galaxy.  However, there are other galaxies 
with dust lanes or morphological asymmetries 
that are not TF outliers, so we do not think that the intrinsic 
scatter of the sample is simply driven by rare ``oddballs''.  

Figure~\ref{fig:iband.V80.deltedb}b shows the impact of removing, in succession, the data 
points with the lowest likelihood (beginning with the seven 
shown in Figure~\ref{fig:iband.V80.deltedb}a), then refitting the TF relation.  Filled
circles show the estimated forward scatter as data points 
are removed, while open circles show the corresponding 
quantity $\sigma'$ $=$ $\sigma_{\rm inv}$/$a_{\rm inv}$ of the inverse
relation.  The inset panel shows the best-fit slopes $a$ 
and $a'$ $=$ 1/$a_{\rm inv}$.  Removing the first data point, 
J021941.13-001520.4, produces a noticeable drop in $\sigma$, 
from 0.42 to 0.40 mag.  Since our estimated inclination 
for this galaxy could be inaccurate for the reasons mentioned 
above, we think there is a reasonable case for lowering our 
estimated values of the intrinsic TF scatter by $\sim$ 0.02 mag 
(less than our {\it statistical} uncertainty of $\sim$ 0.035 mag).  
However, removing subsequent data points produces only a steady, 
approximately linear decrease of the estimated intrinsic scatter.  
If the TF residuals were drawn from a Gaussian of width 0.4 mag, 
then the estimated scatter should show an approximately linear 
decrease to zero at N$_{\rm removed}$ $=$ 160 (a claim we have
tested with Monte Carlo experiments).  The fact that our estimated
scatter goes to zero at N$_{\rm removed}$ $\approx$ 90 suggests 
that we may have overestimated the observational 
errors.  If we set the internal extinction correction and peculiar 
velocity uncertainties to zero and repeat the above procedure, 
the estimated scatter of the forward TF relation approaches 
zero at N$_{\rm removed}$ $\sim$ 130.

\begin{figure}
\epsscale{0.8}
\plotone{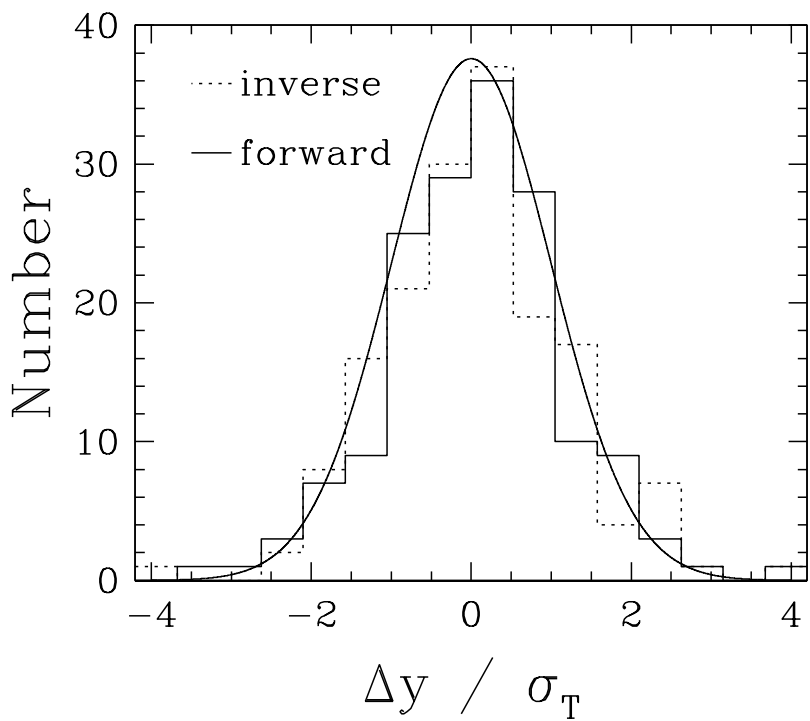}
\caption{Distribution of residuals for the $i$-band TF relation,
with $\Delta y \equiv y-\bar{y}(x)$, 
$y \equiv M_i$ for the forward (solid histogram)
relation and $y\equiv \eta$ for the inverse
(dotted histogram) relation, and the total scatter
for a galaxy with observational errors $\sigma_x$,
$\sigma_y$ defined by 
$\sigma_T^2$ = $\sigma^2$ + 
$\sigma_y^2$ + $(a \sigma_x)^2$. 
The smooth solid curve is a Gaussian of unit dispersion.
The TF outlier J021941.13-001520.4
 for both the forward and inverse relations is at 
$\Delta y$/$\sigma_T$ $\sim$ $-4$.}
\label{fig:deltay}
\end{figure}

Closely related to the role of outliers is the question 
of the residual distribution. Figure~\ref{fig:deltay} plots the histograms
of $\Delta$$y$/$\sigma_T$, where $\Delta$$y$ is the 
deviation of each galaxy from the mean forward (solid) or
inverse (dotted) relation and $\sigma_T$= $[ \sigma^2 + 
\sigma_y + a^2 \sigma_x^2]^{1/2}$ is the quadrature sum of 
the intrinsic scatter and the galaxy's observational error.  
With the exception of the one largest outlier in the 
forward and inverse relations, both histograms are approximately Gaussian in 
form.  (The largest outlier galaxy has $\Delta$$y$/$\sigma_T$ 
$=$ $-4.2$.)  This is further evidence that the intrinsic scatter 
of our sample is not driven by rare outliers. 

Although the differing slopes of the forward and inverse 
relation (specifically, the fact that $a$ $\neq$ 1/$a_{\rm inv}$) 
makes them appear superficially different, Figure~\ref{fig:deltay} shows
that they both lead to nearly Gaussian residual distributions.
The forward
and inverse relations with Gaussian intrinsic scatter are 
equally good, and essentially equivalent, descriptions of 
the {\it two-dimensional} distribution of galaxies in 
the ($M_{\lambda}$,$\eta$) plane, over the range covered by
our sample.  Theoretical models of the galaxy population 
should explain this full distribution, not just a slope 
and intercept whose values necessarily depend on the fitting
procedure.  Similarly, there is no particular virtue to ``orthogonal'' 
fitting procedures that treat the two observables symmetrically 
-- they would have intermediate slopes, but they would 
presumably lead to a similar two-dimensional distribution 
once the scatter was properly accounted for.  

\begin{figure}
\plotone{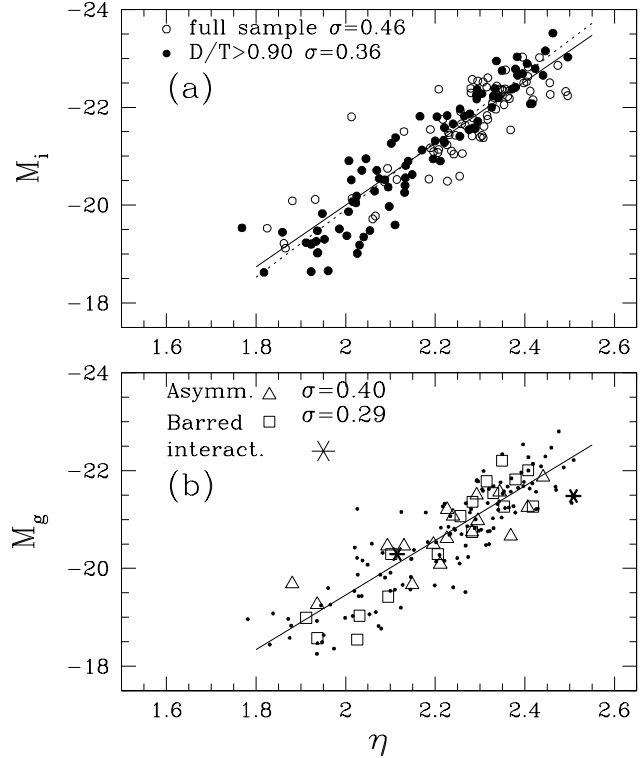}
\caption{
TF relations for morphologically identified subsamples.
(Top) Filled circles represent galaxies with disk-to-total
flux ratios $D/T>0.9$; open circles represent other
sample galaxies.  Solid and dotted lines show best-fit $i$-band
TF relations for the full and disk-dominated samples,
respectively.
(Bottom) TF relation in $g$-band with asymmetrical,
barred, and ``interacting'' galaxies (with nearby neighbors)
marked by triangles, squares, and asterisks, respectively.
The solid line is the best-fit relation for the full data set.
}
\label{fig:prunned}
\end{figure}

Our estimate of $\sigma = 0.42$ mag (for the 
\flagabc\ $i$-band forward relation) is larger than 
at least some previous estimates of the TF intrinsic 
scatter.  It is interesting to explore whether this is a 
consequence of our broader morphological selection.  
Figure~\ref{fig:prunned} compares the $i$-band TF relation of our full sample 
to a disk-dominated subset that have, according to the
GALFIT bulge-disk decomposition, $i$-band disk-to-total flux 
fractions $D/T$ $\geq$ 0.9 (see P05 for an extensive 
discussion of this subset).  The intrinsic scatter drops 
to $\sigma$ $=$ 0.36 mag, a statistically significant 
but modest decrease.  Selecting disk-dominated systems
does not yield the small values of the intrinsic scatter found in 
some previous studies.  We do not see any clear evidence 
for a luminosity dependence of the intrinsic scatter \citep{gio97}, 
but our sample is relatively small for detecting such an effect.

Figure~\ref{fig:prunned}b shows the $g$-band TF relation of the 
full sample, with barred, asymmetric, and 
``interacting'' galaxies marked separately.  Barred galaxies are identified
by visual inspection of the $i$-band images, since 
bars are usually red.  We find 17 galaxies with discernible bars, 
but this is an under-estimate of the true barred fraction 
because bars are difficult to detect in highly inclined systems.  
The asymmetric galaxies are selected by 
visual inspection of the $g$-band images, 
because warps and subtle interactions 
cause a boost in star formation and bluer colors.  
We find 27 galaxies with asymmetries.  The two 
``interacting'' galaxies have close companions.
Figure~\ref{fig:prunned}b shows no clear evidence for 
any of these three classes of galaxies to be systematically
offset from the full TF relation slope, intercept, or 
scatter.  Separate TF fits to the barred and asymmetric 
subsets yield intrinsic scatter estimates of 
$\sigma$ $=$ 0.29 $\pm$ 0.09 and 0.40 $\pm$ 0.08, respectively.  
We conclude that the slightly larger scatter of the full 
sample is not driven by these morphologically 
distinct populations.  

\subsection{Comparison to Previous Studies} 
\begin{figure}
\epsscale{1.0}
\plotone{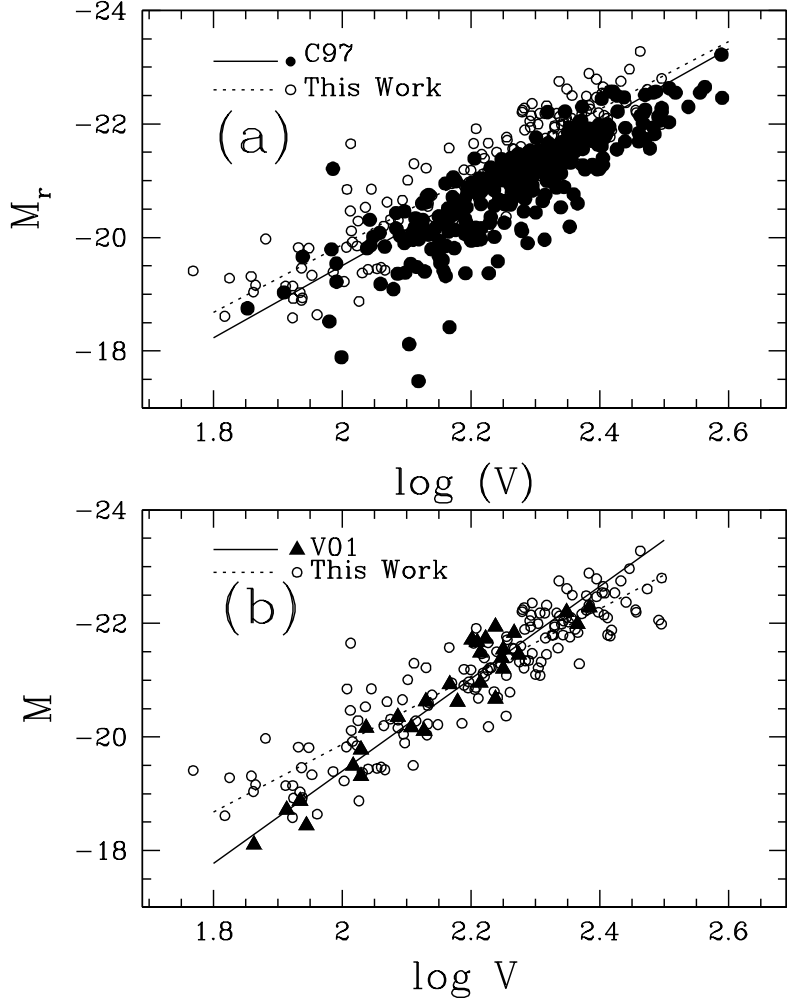}
\caption{
Comparison to results from C97's field sample and V01's sample
from the Ursa Major cluster.
(Top) Filled and open circles show galaxies from C97 and 
from this paper, respectively; dotted and solid curves show the 
corresponding forward TF fits.
Velocities are $\Veight$ for our points and $V_{2.2}$ for C97.  
(Bottom) Same as above, but with filled triangles 
representing the V01 data.  Absolute magnitudes are 
SDSS $r$ for our points and Johnson $R$ for V01.  
Velocities for V01 are $\Vflat$,
measured from the flat portion of HI synthesis rotation curves.
}
\label{fig:otherworks}
\end{figure}

Figure~\ref{fig:otherworks} compares our TF results to those of two previous 
studies:  C97, whose analysis techniques are similar 
to our own, and V01, who presents a comprehensive investigation 
of spirals in the Ursa Major cluster.  C97's sample, 
shown in Figure~\ref{fig:otherworks}a, consists of 304 Sb-Sc galaxies 
selected from the UGC.  For his data points, 
we use his preferred definition of rotation speed 
as the amplitude of the arc-tangent function fit to 
the \halpha\ rotation curve at 2.2 
disk scale lengths, while for our data points we 
show $\Veight$.  We ignore slight differences between 
the Gunn $r$-band and the SDSS $r$-band.  Relative to 
our sample, C97's is much more strongly weighted 
towards luminous galaxies.  Despite these differences, 
the TF relations agree well (solid and dashed lines), 
with our slope $a$ $=$ $-5.96$ $\pm$ 0.20 slightly 
shallower than C97's $a$ $=$ $-6.36$ $\pm$ 0.22 (C97, Table 4).  
(If we fit the C97 data with our routines, we obtain a similar 
result.)

Figure~\ref{fig:otherworks}b shows the $R$-band TF relation for 29 Ursa Major 
cluster galaxies from V01, with circular velocities defined
from the flat portion of rotation curves measured by HI synthesis
imaging, compared to the $r$-band TF relation measured here.
The V01 TF relation is clearly steeper; most of the difference
is attributable to the low luminosity galaxies, which rotate
faster at fixed $M_R$, though the clump of galaxies at
$M_R \approx -21.5$ is also rotating slightly slower.
The obvious potential culprit is the velocity width definition,
with HI rotation curves yielding circular velocities that are
systematically higher than $\Veight$ for low luminosity galaxies.
However, using the trends found for the subset of C97 
galaxies with both \halpha\ and HI velocity widths,
P05 conclude that this difference can only account for about
half of the difference in the TF slopes.
Since the number of low luminosity galaxies in V01's sample
is small, the impact of this difference in velocity definition
could perhaps be enhanced by small number statistics.
Alternatively, the finite depth of the Ursa Major cluster
could contribute to a steeper slope if the fainter galaxies
happen to lie preferentially on the far side, but this is un-likely.
The bias-corrected $r$-band slope, discussed in Appendix B, 
steepens to a value  ($-7.02$) much closer to the V01 $R$-band TF sample.  
Therefore, one can conclude that the major differences are a combination 
of the velocity width definition and sample selection in this study.
A larger sample of systems with
both H$\alpha$ and HI synthesis rotation curves would
help shed light on the origin of this difference.

In Table~\ref{tbl:TFotherworks} we present the slopes and 
intercepts for several TF relations commonly used to 
compare theoretical predictions to observations.  The intercepts
are not listed due to variations in the assumed value of $H_0$, 
and different internal extinction corrections.  
The difference between slopes, in a given band, is several 
times the typical statistical uncertainties, with no obvious
trends.  As shown above, different velocity width measurements  
can account for some of those differences.  K02 is also able to 
measure slopes for the MAT and CF field galaxy samples that are similar to 
the K02 slopes after applying a consistent internal extinction 
correction, velocity width definition, and fitting method.  
The intrinsic scatter is typically smaller in the studies using 
HI velocity widths.  Those samples are primarily used
as distance indicators to study peculiar velocity fields, and  
represent more pruned samples.  For example, the small
scatter measured by \cite{pt92} was measured for a 
pruned sample of nearby spiral galaxies having  
well measured distances.   In samples with optical velocity 
widths and careful considerations of biases, the intrinsic 
scatter appears to be 0.3-0.5 magnitudes.  
Given our sample's broad selection criteria in terms of galaxy 
Hubble type, well defined magnitude and redshift cuts, 
and further distances we consider our TF relation to suffer 
less peculiar velocity uncertainties and Hubble type selection 
effects.  However, the large differences in the slopes 
may be due to the Malquist-type biases discussed in Appendix B.

\section{Residual Correlations}
To understand the physical sources 
of scatter in the TF relation, we want to 
investigate the correlation of TF residuals 
with other galaxy properties.  For 
example, in the case of ellipticals, 
the correlation of residuals from the Faber-Jackson relation 
\citep{fab76} with galaxy 
size led to the recognition that 
ellipticals occupy a fundamental plane that 
largely corresponds to the virial relation 
for the stellar component \citep{djo87,dre87}.  
The lack of a similar correlation for disks 
implies that disk gravity does not dominate 
the rotation speed at radii used for TF 
investigations \citep{cou99,piz05}.  If 
the rotation velocity of disk galaxies is more fundamentally 
related to the stellar mass than 
to the luminosity, there should be a correlation of TF 
residual with color, which tracks the 
stellar mass-to-light ratio (K02).  For a 
theoretical discussion of some of these points, 
see \cite{con01}, \cite{she02}, \cite{dut06}, and \cite{gne06}.  

\begin{figure}
\plotone{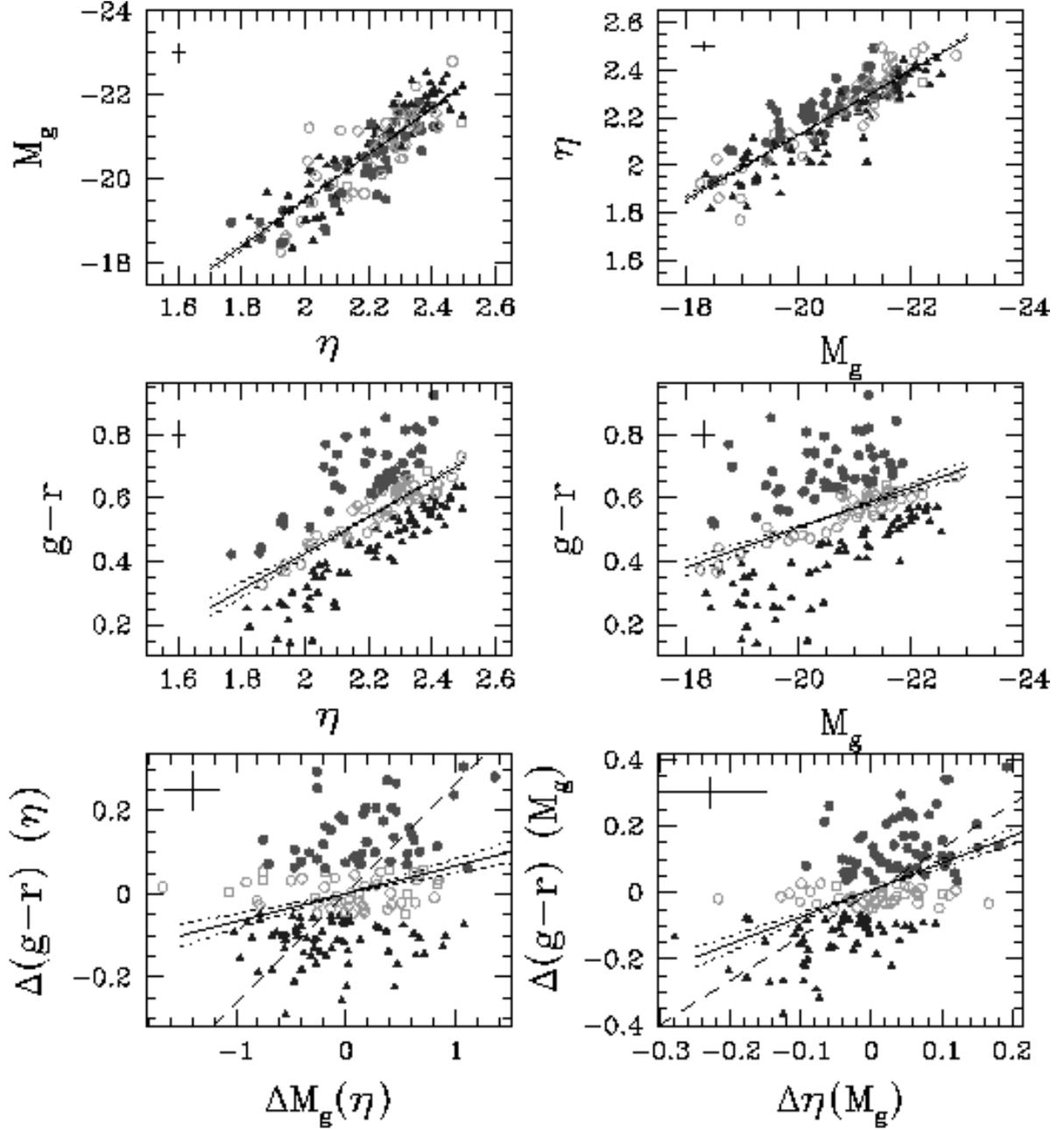}
\caption{Correlation of $g$-band TF residuals with $g-r$ color residuals,
for the forward (left) and inverse (right) relations.
In the middle panels, solid lines show the best-fit mean
trend of (extinction corrected) $g-r$ with $\eta$ or $M_g$, with dotted curves
showing the bootstrap uncertainties.  Filled circles, open circles,
and triangles show the reddest, intermediate, and bluest 1/3
of the galaxies relative to these mean relations; the same
point types are used for each galaxy in the upper and lower panels.
Bottom panels plot residual from the mean color relation
against TF residual.  Solid and dotted lines show the best-fit
residual correlation and $1\sigma$ uncertainties.
Dashed lines show the expected slope for pure self-gravitating
disks with mass-to-light ratio as a function of color
according to \cite{bel03}.
Error crosses in each panel show median values of 
the observational errors.
}
\label{fig:gcolorres}
\end{figure}

\begin{figure}
\plotone{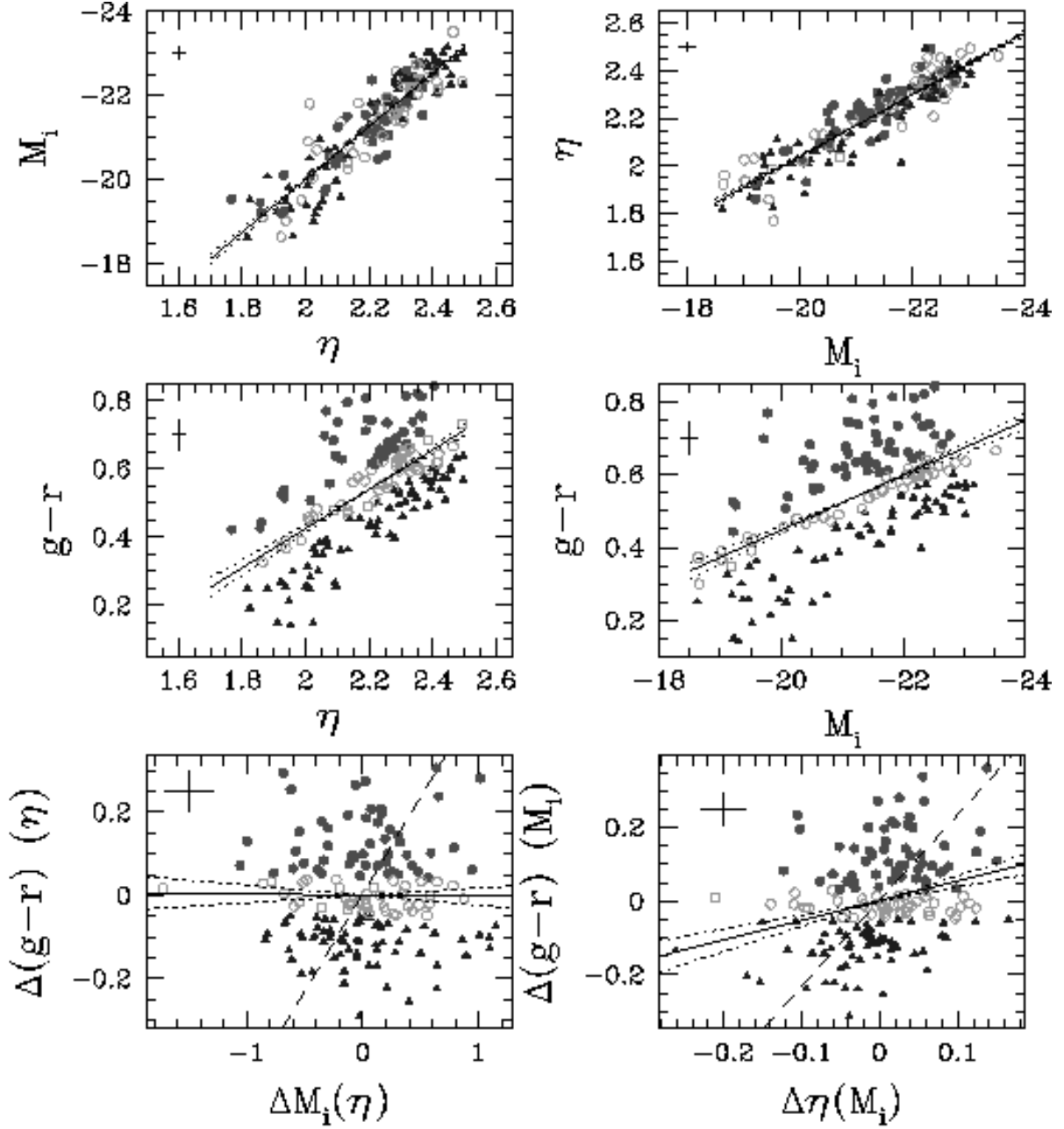}
\caption{Correlation of $i$-band TF residual with residual from
the mean relation between $i$-band half-light radius $R_i$ and
$M_i$ or $\eta$.  Format is similar to Fig.~\ref{fig:gcolorres}.
}
\label{fig:icolorres}
\end{figure}

The top panels of Figure~\ref{fig:gcolorres} show the forward 
and inverse $g$-band TF relations.  The middle 
panels show the correlation of the extinction 
corrected $g$-$r$ color with $\eta$ and $M_{\lambda}$.  
The lines show the maximum likelihood best-fit 
relations, with parameters and bootstrap uncertainties 
listed in Table~\ref{tbl:res}.  Point types encode the residual 
color relative to this mean relation, with filled 
circles, open circles, and triangles showing 
the reddest, intermediate, and bluest 1/3 of the
galaxies.  The same point type is used for each 
galaxy in the upper panels, and one can see that 
red galaxies tend to be slightly underluminous 
in the forward relation.  The top right 
panel of Figure~\ref{fig:gcolorres} clearly shows that red
galaxies tend to rotate faster at fixed $\Mg$.
  The bottom panels plot the residual from the 
TF relation against the residual from the 
color-$\eta$ or color-$\Mg$ relations; solid lines
show the maximum likelihood fit to the mean correlation 
of residuals, and dotted lines show bootstrap 
uncertainties.  The residuals are correlated, again 
more clearly for the inverse relation, 
but there is substantial 
scatter that is large compared to the observational errors.  
Note that while the bootstrap errors on the best-fit linear slopes
are small, the data are not well described by any linear
relation with zero intrinsic scatter, and the derived slopes 
therefore depend significantly on the fitting procedure.    
 Figure~\ref{fig:icolorres} shows similar results for the $i$-band TF relation.  
The correlations between color and $i$-band TF residuals 
are much weaker, being essentially absent in the 
forward relation.  

In the population synthesis modeling of \cite{bel03},  
the mass-to-light ratio of a stellar population changes 
with $g$-$r$ color roughly as  
$M_*/L_g$ $\propto$ $(g-r)^{1.52}$ in the $g$-band, and 
$M_*/L_i$ $\propto$ $(g-r)^{0.86}$ in the $i$-band.  For pure 
self-gravitating disks, the circular velocity should 
correlate with the stellar mass as $V^2$ $\propto$ $M_*$ 
at fixed scale length.  Variations of the stellar mass-to-light
ratio would therefore produce inverse TF residual 
correlations of the form  $\Delta (g-r)$ $\simeq$ 
2$\Delta \eta (\Mg) /1.52$ and $\Delta (g-r)$ $\simeq$ 
2$\Delta \eta (\Mi) /0.86$, and forward TF residuals 
of the form $\Delta (g-r)$ $\simeq$ $-0.4\Delta \Mg (\eta) 
/ 1.52 $ and $\Delta (g-r)$ $\simeq$ $-0.4\Delta \Mi (\eta) 
/0.86 $, in the absence of other effects.  Dashed 
lines in Figure~\ref{fig:gcolorres} and Figure~\ref{fig:icolorres}
show the correlation slopes predicted by this simplistic 
model.  

The residual correlations in the bottom panels of 
Figure~\ref{fig:gcolorres} and Figure~\ref{fig:icolorres} have
the correct sign expected for variations of $M_*/L$ 
with stellar populations.  Therefore, we concur with the 
conclusion of K02 based on the NFGS, that these 
variations account for some of the scatter in the TF 
relation.  However, it is clear from the scatter about 
the mean residual correlation, especially in the $i$-band, that 
these variations do not account for {\it much} of 
the intrinsic scatter.  We have tried the experiment 
of changing $\eta$ for each galaxy by an amount $-\Delta \eta$ 
predicted from its $g$-$r$ color using the best-fit 
slope to the inverse TF residual correlations from 
Figure~\ref{fig:gcolorres} and Figure~\ref{fig:icolorres}, 
then refitting the TF relation.  
This procedure reduces the estimated intrinsic scatter 
of the inverse relation from $\sigma'$ $=$   0.073 to 0.057 
in the $g$-band and from $\sigma'$ $=$ 0.061 to 0.057 
in the $i$-band, drops of 22\% and 7\%, respectively.

We can also explore the correlations of TF residuals 
with structural parameters.  
Figure~\ref{fig:radres} presents the correlation of TF residuals 
with $i$-band half-light radius $R_i$, determined 
from the GALFIT model fits to the $i$-band images, 
in the same format as 
Figure~\ref{fig:gcolorres} and Figure~\ref{fig:icolorres}.  
In the forward relation, there is a slight tendency 
for larger disks at fixed $\eta$ to be slightly 
more luminous (upper left panel).  This trend 
leads to a weak correlation between the TF residual 
and the residual from the mean $R_i$-$\eta$ relation, 
though the scatter is large compared to the mean 
correlation (and to the observational errors).  The inverse 
fits reveal no trend of residual $\Veight$\ with residual 
$R_i$ at fixed $\Mi$.  Dashed lines show the predictions 
of a pure self-gravitating disk model, with $V^2$ 
$\propto$ 1/$R_i$ at fixed $\Mi$.  

These weak or absent residual-radius correlations 
confirm the results of \cite{cou99}, now with a 
more broadly selected sample and smaller observational 
errors per galaxy.  We found similar results for a 
disk-dominated subset in P05, 
using $V_{2.2}$ as a rotation measure.  For the 
full sample, we have also investigated using the 
velocity at radii containing 60\% to 90\% of the 
total $i$-band flux, again with similar 
results.  As discussed by \cite{cou99}, and in greater detail by 
\cite{dut06} and \cite{gne06}, the absence of strong 
radius residuals imposes strong constraints on the 
contribution of disk gravity to the rotation speed; 
disk gravity should cause more compact galaxies 
to rotate faster, and the impact of the disk on the 
inner halo profile \citep{blu86,gne04} should 
amplify this effect.  Explaining the observed lack 
of correlation requires ``sub-maximal'' disks, and it is 
more easily accomplished if disks do not produce 
adiabatic contraction of halos (see \citealt{dut06,cou99,gne06}).  

\begin{figure}
\plotone{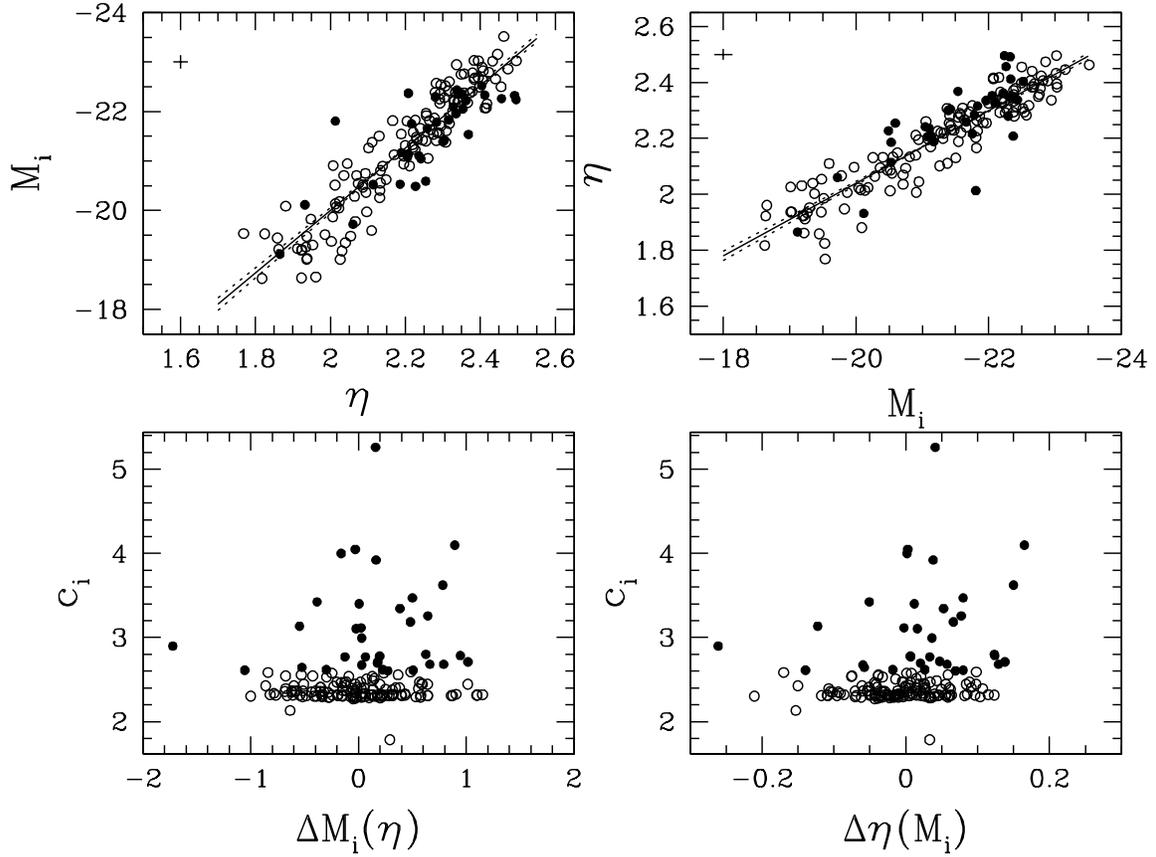}
\caption{Correlation of $i$-band TF residual with $i$-band concentration
index $c_i = r_{90}/r_{50}$.  Lower panels plot $c_i$ vs. TF
residual.  In all panels, filled and open circles show ``early''
and ``late'' galaxy types, respectively, with the division
between types at $c_i=2.6$.
}
\label{fig:conres}
\end{figure}

Figure~\ref{fig:conres} examines the dependence of the 
TF residuals on morphology, as quantified by the 
concentration index $c_i$ $=$ $R_{90}/R_{50}$, 
where $R_{90}$ and $R_{50}$ are the radii 
enclosing 90\% and 50\% of the $i$-band 
Petrosian flux, as determined by the 
SDSS photometric pipelines .  An index
of $c$ $\sim$ 2.6 is often adopted as the separation 
between early-type (high concentration) and
late-type galaxies (e.g., \citealt{str01}).  We 
plot early-type and late-type galaxies as filled
and open circles in the top panels of Figure~\ref{fig:conres}, 
using the $c_i$ $=$ 2.6 division, and the bottom 
panels plot $c_i$ against the TF residuals.  On 
average, early-type galaxies tend to be fainter 
at fixed $\eta$ or rotating faster at fixed 
$\Mi$, though the mean offset is small compared to 
the scatter.  The mean residuals for early-type 
galaxies are $<\Delta \Mi>$ $=$ 0.14$\pm$0.09 and
$<\Delta \eta>$ $=$ 0.026$\pm$0.015, rising to 
0.19$\pm$0.08 and 0.034$\pm$0.012 if we exclude the 
single strong outlier (J021941.13-001520.4) discussed in 
\S 5.3.  This trend agrees with the finding by K02 
that Sa galaxies are fainter than later type 
spirals at fixed rotation speed ( see also G97 and \cite{mas06}). The sign of 
the observed trend agrees with expectations from 
older, redder stellar populations of early-type galaxies.  
However the separation in Figure~\ref{fig:conres}, while modest, 
appears somewhat clearer than the trend with color
seen in Figure~\ref{fig:icolorres}, suggesting a direct correlation 
of rotation speed with morphological structure.  
Using the GALFIT D/T ratio, instead of $c_i$, gives
similar results. 

\section{Conclusion}
We have measured the TF relation for a sample 
of 162 galaxies selected from the SDSS, using 
follow-up spectroscopy to obtain \halpha\ rotation
curves.  We targeted 234 galaxies from the SDSS 
spectroscopic galaxy catalog that have a roughly  
flat absolute magnitude distribution in the 
range $-18.5 > \Mr > -22$.  For the purpose 
of testing galaxy formation models, our sample 
has several advantages relative to most previous 
TF studies.  Our target selection is blind to morphology, except for an 
$i$-band axis ratio cut of 0.6.
Our completeness is high, and the galaxies with
usable \halpha\ rotation curves have distributions
of color and concentration similar to a large control sample
of galaxies selected from DR4 with the same $\Mr$ distribution
but no inclination cut or \halpha\ requirement.
Due to the uniform SDSS photometry and the selection 
of galaxies at redshifts greater than 5000 $\kms$, 
all sources of observational error are small compared to the 
estimated intrinsic scatter.  The uniform multi-wavelength SDSS photometry 
allows us to investigate the TF relations and the correlations between 
TF residuals, color residuals, and other structural 
parameters in the $g$, $r$, $i$, and $z$-bands with 
minimal photometric uncertainties.  

We adopt $\Veight$\ as a measure of the rotation 
speed and use extinction corrected absolute 
magnitudes.   
We estimate the slope, intercept, and intrinsic scatter
of the TF relations simultaneously, using a maximum
likelihood procedure that accounts for individual
observational errors in absolute magnitude and rotation speed.
We measure forward TF slopes between 
$-5.5$ and $-6.6\,{\rm mag}/{\rm log_{10}}(\kms)$, 
with typical uncertainty of 0.2, and
intercepts between $-20.7$ and $-21.5$ mag at $\eta = 2.22$, 
with typical uncertainty of 0.04.
The slope becomes systematically steeper with wavelength.
Inverse fits always yield steeper slopes, but once the effects
of Gaussian scatter are included, the forward and inverse
fits describe essentially the same two-dimensional
distribution of data points over the range covered by our data.
Corrections to these slopes due to Malmquist-type biases
are discussed in Appendix B, where it is shown that our
observed TF relation can be reproduced with a steeper 
slope and larger intrinsic scatter given certain assumptions 
about selection effects.

The intrinsic scatter appears to be nearly independent 
of wavelength or fitting procedure, typically $\sigma = 0.42-0.46$ mag,
with a higher value (0.54 mag) for the inverse fit in $g$-band.
The distribution of residuals is approximately Gaussian
for both the forward and inverse relations, and there is no
indication of rare outliers inflating the TF scatter,
with the possible exception of one galaxy whose axis
ratio may be incorrectly measured because of spiral arms.
Omitting this one system reduces the estimated intrinsic
scatter by 0.02 mag.
The intrinsic scatter is slightly smaller 
for a disk-dominated subset of galaxies, decreasing 
in the $i$-band from 0.42 mag to 0.36 mag for galaxies 
having disk-to-total flux ratios greater than 0.9.  
Morphologically asymmetric, barred, or possibly 
interacting galaxies show no clear evidence for offsets
from the mean TF relation or for larger scatter,
though our statistics for these subsets are limited.

In an attempt to understand the origin of scatter in the TF relation,
we study correlations between TF residuals and other 
galaxy properties: $g-r$ color, half-light radius,
and $i$-band concentration index.   
The $g$-band TF residual shows a significant correlation
with color, in the sense that bluer galaxies are brighter
at fixed $\eta$ or (more clearly visible in the data) that redder
galaxies rotate faster at fixed $M_g$.  However, these
correlations are much weaker in the $i$-band, and they are
much weaker than predicted for pure self-gravitating
disks with the stellar mass-to-light ratios predicted
from $g-r$ color.  The correlation with half-light
radius is very weak, with a slight tendency for larger
galaxies to be more luminous at fixed $\eta$.
More concentrated (earlier type) galaxies tend to be
slightly fainter at fixed $\eta$ or to rotate slightly
faster at fixed $M_i$, but the trend is weak compared
to the TF scatter.

Our TF relation is similar to that found for field spirals
by C97, who used similar analysis methods.
The TF relation found by V01 for Ursa Major spirals, using
the flat portion of HI rotation curves as a velocity measure,
is significantly steeper than ours.  This difference may
partly reflect the difference in velocity width definitions, 
and Malmquist-type biases as described in 
Appendix B.  The residual correlations in our sample are
similar to those found by K02 using the Nearby Field Galaxy Survey
of \cite{jan00}, but we find that stellar population
effects explain only a small fraction of the scatter 
in the TF relation.  

The nearly constant intrinsic scatter of the TF relation has 
interesting implications for galaxy formation 
theories.  The scatter is only slightly larger in bluer bands 
and only slightly smaller for a morphologically selected disk-dominated
sub-sample of galaxies, and the correlation of TF residuals
with other properties explains only a small fraction of the 
scatter.  The lack of correlation with radius implies that 
disk gravity has small influence on galaxy dynamics at
the radii used for TF measurements; 
this observational result provides a strong constraint 
on galaxy formation models \citep{dut06,gne06}.  The 
weak correlation with color, size, and morphology 
implies that most of the scatter in the TF relation
arises from genuine variation in the ratio 
of dark matter to baryonic matter from halo to halo.  

Our results suggest many fruitful directions for future
observational investigations, each of them comparable in
magnitude to the one undertaken here.  
Absorption-line velocity measurements, which require
longer exposures than our \halpha\ emission line measurements,
could yield rotation velocities for the remaining 30\%
of our sample, showing whether galaxies without \halpha\ 
emission follow a systematically different TF relation.
HI measurements would allow detailed comparison between
TF relations defined by \halpha\ and HI velocity widths
(see C97) for our broadly selected galaxy sample, and they
would provide gas masses and gas fractions as additional
parameters for residual correlation studies, which would
be valuable for testing theoretical models
\citep{dut06,gne06}.
Extending the TF relation to brighter and fainter luminosities 
is of great interest (e.g., \citealt{mcg00}),
though complications arise from the growing paucity of
disks in the first case and the increasing importance
of non-circular motions in the second.
Finally, it is clear that detailed investigation of
TF residuals would benefit from a much larger sample
with similarly broad selection criteria and similarly
small observational errors, so that one could, for example,
investigate residual correlations with size, morphology,
or environment within restricted regions of luminosity-color space.
Since large samples of galaxies ($\ge 2000$) have been observed for TF
studies in the past (see, e.g., \citealt{cou03,mas06} and
references therein), it may be possible to construct
such a sample largely from archival data, if the data
quality can be made sufficiently homogeneous and the effects
of sample selection can be sufficiently well understood.
Our data already provide a well defined and highly
constraining target for theoretical models of galaxy
formation.  Extended observational studies would improve
our ability to deduce the essential physics of disk
galaxy formation empirically.

\acknowledgments

We thank Vijay Narayanan, Jeff Munn, and Robert Lupton
for assistance in the early stages of this project. We thank  
Oleg Gnedin, Aaron Dutton, Susan Kassin, Vladimir Avila-Reese, 
and Sheila Kannappan for helpful discussions and comments.
We also thank the anonymous referee for useful suggestions
and comments.  JP and DHW acknowledge support from NSF Grants AST-0098584 and 
AST-0407125.  We thank the staff at MDM and Calar Alto 
observatories for support.

Funding for the SDSS and SDSS-II has been provided by the 
Alfred P. Sloan Foundation, the Participating Institutions, 
the National Science Foundation, the U.S. Department of Energy, 
the National Aeronautics and Space Administration, the Japanese 
Monbukagakusho, the Max Planck Society, and the Higher 
Education Funding Council for England. 
The SDSS Web Site is http://www.sdss.org/.

The SDSS is managed by the Astrophysical Research Consortium 
for the Participating Institutions. The Participating 
Institutions are the American Museum of Natural History, 
Astrophysical Institute Potsdam, University of Basel, Cambridge 
University, Case Western Reserve University, University of 
Chicago, Drexel University, Fermilab, the Institute for Advanced 
Study, the Japan Participation Group, Johns Hopkins University, 
the Joint Institute for Nuclear Astrophysics, the Kavli Institute 
for Particle Astrophysics and Cosmology, the Korean Scientist 
Group, the Chinese Academy of Sciences (LAMOST), Los Alamos 
National Laboratory, the Max-Planck-Institute for Astronomy 
(MPIA), the Max-Planck-Institute for Astrophysics (MPA), New 
Mexico State University, Ohio State University, University of 
Pittsburgh, University of Portsmouth, Princeton University, the 
United States Naval Observatory, and the University of Washington.

This paper is based in part on observations obtained in the framework
of the Calar Alto Key Project for SDSS Follow-up Observations (\citealt{gre03}) 
at the German-Spanish Astronomical Centre, Calar Alto Observatory, 
operated by the Max Planck Institute for Astronomy, Heidelberg, jointly 
with the Spanish National Commission for Astronomy.

\clearpage
\appendix

\section{TF Parameters for Alternative Parameters and Sample Cuts}

Our standard TF fits, presented in Table~\ref{tbl:tf},
use all \flagabc\ galaxies, absolute magnitudes corrected
for internal extinction, velocities defined by $\Veight$, and
equal contribution of each galaxy
to the likelihood sums of equation~(\ref{eqn:like}).
Table~\ref{tbl:appx} presents TF fits using a number of
variations on parameter definitions and sample cuts or weights.
First, we list results for the full sample with the usual
definitions but weights chosen to approximate a truly flat
absolute magnitude distribution in the range $-18.5 > M_r > -22$
(see \S 5.2).  Next, we list results for the
full sample with the usual weights but no internal extinction
correction (see Fig.~\ref{fig:gcompare}).
Then we list results for two alternative velocity width definitions:
the value of the arc-tangent functional fit evaluated at 2.2
disk scale lengths as determined from the $i$-band GALFIT model ($V_{2.2}$), 
and at the location of the most distant \halpha\ data point ($V_{end}$).
Finally, we list the TF fits for the standard definitions
and \flagab\ galaxies only.
Although slopes and intercepts change somewhat with these
choices, none of our conclusions about the TF relation
would be qualitatively different for any of these 
alternative choices.

\section{Malmquist-type Biases in Our TF Sample.} 
Applying magnitude cuts to TF samples can lead to biased 
estimates of the TF parameters because there are low and
high luminosity galaxies, at fixed velocity, scattered above
or below the magnitude limits.  This can be due to both the intrinsic
scatter and peculiar velocities scattering galaxies 
across selection boundaries.  Most discussions of these
``Malmquist''-type biases have focused on apparent magnitude
limited samples, or on the systematic bias in the derived
peculiar velocity field (e.g., \citealt{lyndenbell88,gould93,teerikorpi93,str95}).  
  Since our selection procedure is quite different from those of
most previous TF surveys, we have tested for Malmquist-type
biases with a simple Monte Carlo experiment.

\begin{figure}
\epsscale{1.0}
\plotone{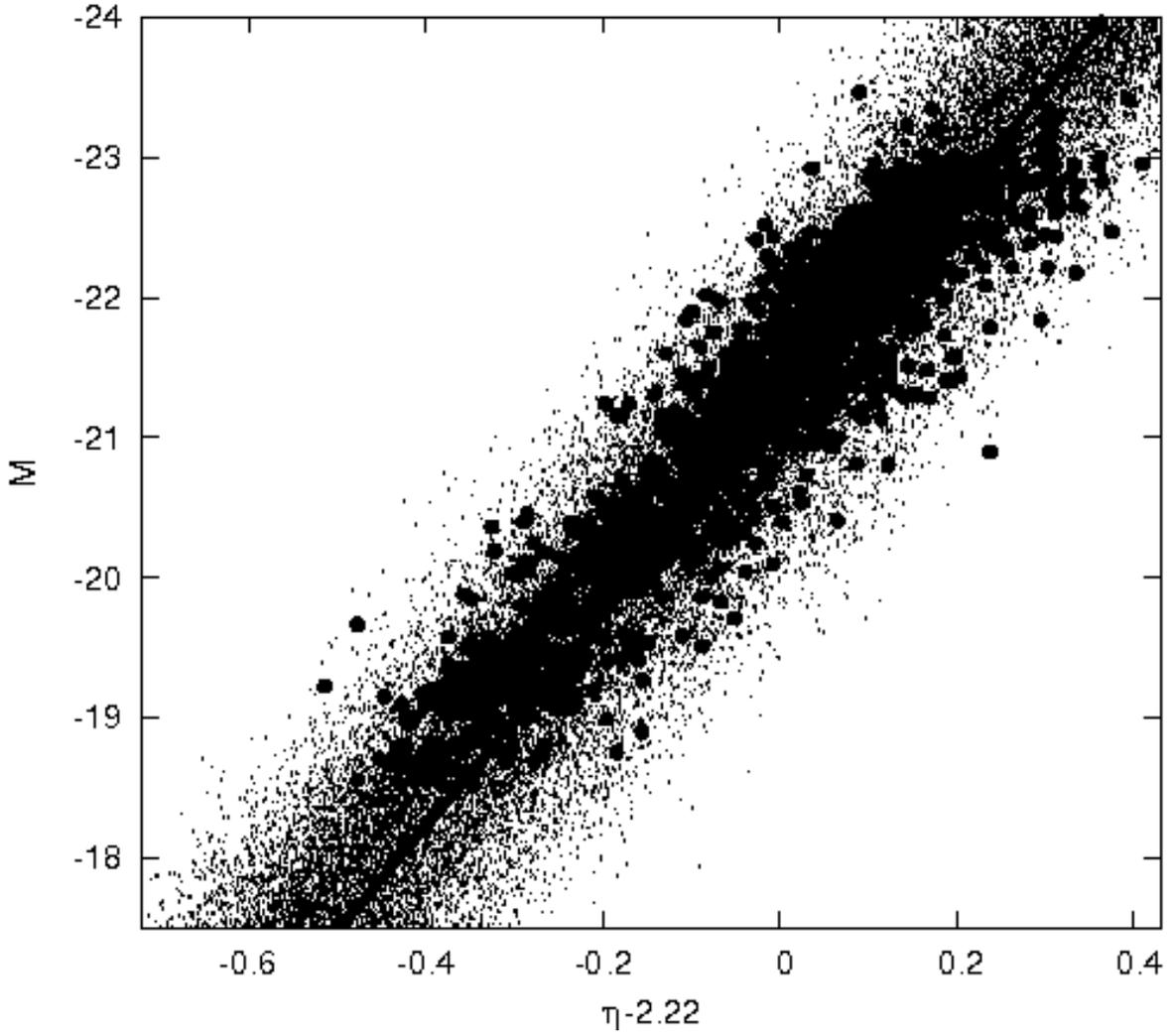}
\caption{ 
A Monte Carlo sample without absolute magnitude cuts (small dots) and 
with our absolute magnitude cuts applied (filled circles).  The relation 
used to generate the sample is shown as a solid line, the dotted line is 
the best-fit to the sample after the cuts have been applied.
}
\label{fig:unbiasTF}
\end{figure}

We first investigate the affect our absolute magnitude-redshift 
cuts and \flagabc\ absolute magnitude distribution will have 
on our measured TF relation.  For our Monte Carlo 
experiment, we generate a sample of $10^5$ galaxies, 
evenly sampled in logarithmic velocity space, 
assigning random positions in a space between 4000 $\kms$ 
and 16000 $\kms$. We then assign absolute magnitudes to each 
galaxy following an assumed TF relation, having a Gaussian 
intrinsic scatter, represented by
\begin{equation}
M = a_{ub} (\eta - \eta_0) + b_{ub} + N(0,\sigma_{ub}),
\end{equation}
where {\it $a_{ub}$}, {\it $b_{ub}$}, and $\sigma_{ub}$ are the 
slope, intercept, and intrinsic scatter for an un-biased power-law
TF relation.  We use $\eta_0 = 2.22$ throughout the procedure.  
We then apply our absolute magnitude-redshift
cuts, outlined in \S 2, and select galaxies according to our 
observed \flagabc\ absolute magnitude distribution, with a 
final Monte Carlo sample that is 10 times larger than our 
\flagabc\ sample, or 1620 galaxies.  Generating a 
Monte Carlo sample with {\it $a_{ub}$} $=$ -7.52,
{\it $b_{ub}$} $=$ -21.23, and $\sigma_{ub}$ $=$ 0.45, 
and applying our cuts, will have a measured TF relation with a 
slope, intercept, and scatter of -6.86, -21.27, 0.33.  This is 
quite a large difference considering our typical statistical uncertainties
on the measure slope, intercept, and scatter being 0.2, 0.04, and 0.04 mag 
respectively.  The effect on the inverse relation is much smaller since the 
cuts are applied to absolute magnitudes.  Figure~\ref{fig:unbiasTF} 
shows this Monte Carlo sample with and without these cuts.
It is clear from  Figure~\ref{fig:unbiasTF}  that the selection criteria 
has excluded high and low luminosity galaxies, shallowing the
slope and reducing the scatter.  

\begin{figure}
\plotone{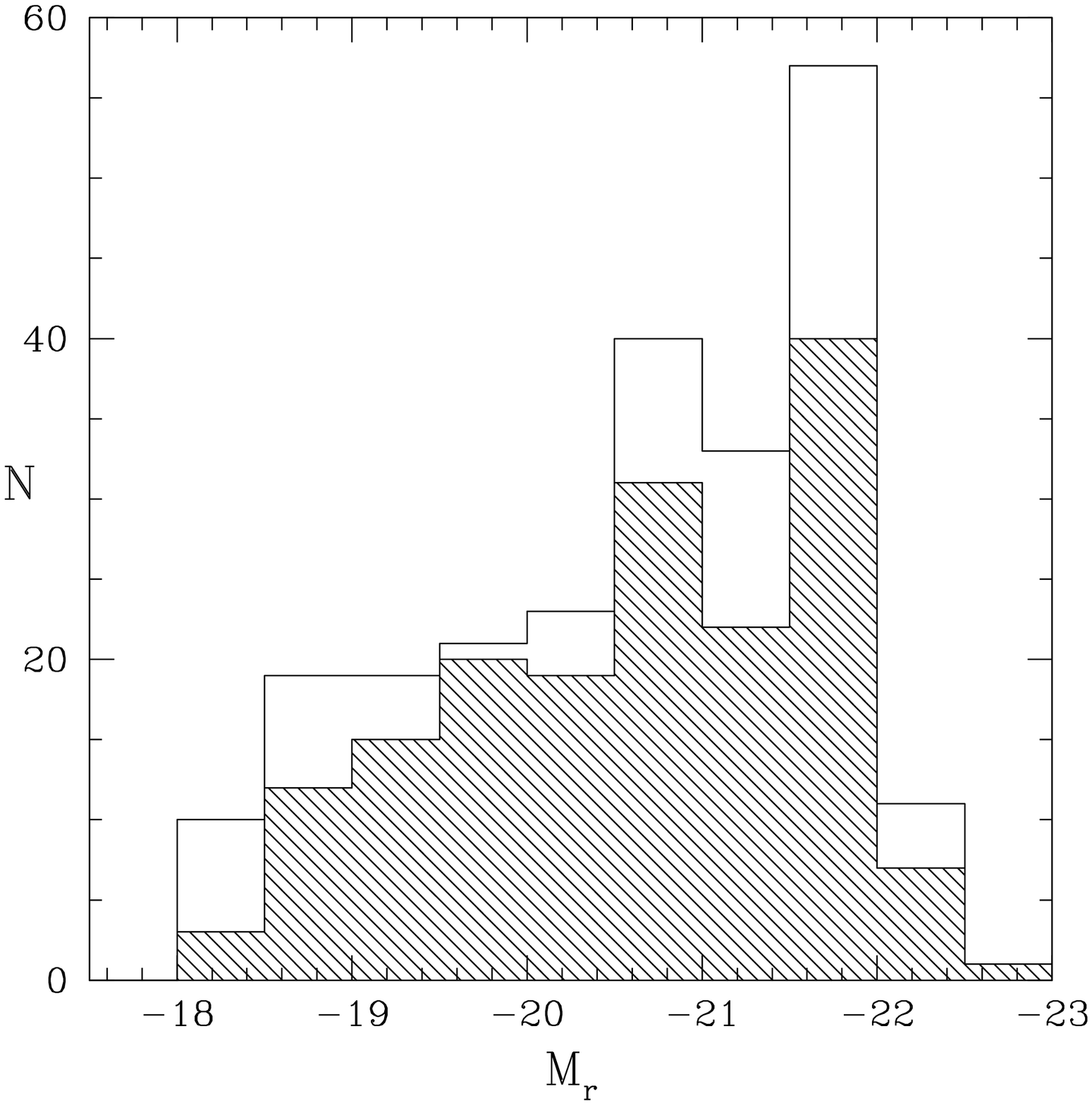}
\caption{Distribution of $r$-band absolute magnitudes for the full sample 
targeted for follow-up \halpha\ observations (234 galaxies, full 
histogram) and the subset of the sample with usable (\flagabc) \halpha\ 
rotation curves (170 galaxies, shaded histogram)} 
\label{fig:histogram}
\end{figure}

The measurement errors in SDSS Petrosian magnitudes are generally
small, and luminosity uncertainties are therefore dominated by
line-of-sight peculiar velocities, which we have assumed in our
analysis to be drawn from a Gaussian of dispersion $300\kms$.
Next we include peculiar velocities in the Monte Carlo sample 
by modifying each galaxy's redshift with a peculiar velocity drawn from
a $300\kms$ Gaussian.  We then apply the same absolute-magnitude
redshift cuts, while matching the \flagabc\ absolute magnitude 
distribution in the shaded area of Figure~\ref{fig:histogram}, then 
measure the TF relation for this sample.  Generating a Monte Carlo sample
with {\it $a_{ub}$} $=$ -7.52,
{\it $b_{ub}$} $=$ -21.23, and $\sigma_{ub}$ $=$ 0.45
 has a measured slope $=$ -6.77, intercept $=$ -21.34, and $\sigma$ $=$ 0.33 mag.  
This is a relatively small, but non-zero, effect.  We include the
300$\kms$ peculiar velocities in the analysis below.  

We now estimate what power-law TF relation and Gaussian 
intrinsic scatter would be required to reproduce the measured
forward TF relations in Table 4, given our selection criteria.
This is done iteratively by varying {\it $a_{ub}$}, {\it $b_{ub}$}, and $\sigma_{ub}$, 
until the fitted parameters, after applying our selection 
criteria and 300$\kms$ peculiar velocities, agree with 
the measure TF relations reported in Table 4.  The forward 
$i$-band TF relation in Table 4 is reproduced with un-biased
values of {\it $a_{ub}$} $=$ -7.35, {\it $b_{ub}$} $=$ -21.27, 
and $\sigma_{ub}$ $=$ 0.58 mag.  Comparing these numbers 
to the best-fit values for the forward $i$-band relation in 
Table 4, shows that the un-biased slope is steeper by $\sim$ 
5 sigma, the intercept changed by several sigma, and the 
un-biased intrinsic scatter is larger by $\sim$ 4-5 sigma. 
Similar results are found for the $g$, $r$, and $z$-bands.  
The un-biased $r$-band TF parameters, that reproduce the
best-fit values in Table 4, are {\it $a_{ub}$} $=$ -7.02, 
{\it $b_{ub}$} $=$ -21.05, and $\sigma_{ub}$ $=$ 0.56 mag.  

Table 4 presents the best-fit TF relation to galaxies 
observed with the selection criteria outlined in \S 2.  
Although the un-biased parameters discussed here are 
quite different from the best-fit TF parameters, the 
best-fit parameters have fewer assumptions than the 
un-biased parameters and have well defined selection 
criteria.  Namely, the un-biased power-law TF relation 
assumes that there is no change in slope at high and low luminosities (contrary to K02), 
and that the best-fit TF parameters in Table 4 are a fit to the  
data and not a Monte Carlo experiment.
However, G97 show that these assumptions have a small
affect on the un-biased parameters.
Therefore, the recommended procedure 
when trying to reproduce the TF relation in the SDSS 
bands would be to apply our selection criteria and 
absolute magnitude distributions to a sample (theoretical 
or observed), then measure the best-fit parameters.


\clearpage

\clearpage

\clearpage

\clearpage

\clearpage

\clearpage

\clearpage

\clearpage

\clearpage

\clearpage

\clearpage

\clearpage

\clearpage

\clearpage
\begin{figure}
\epsscale{0.8}
\plotone{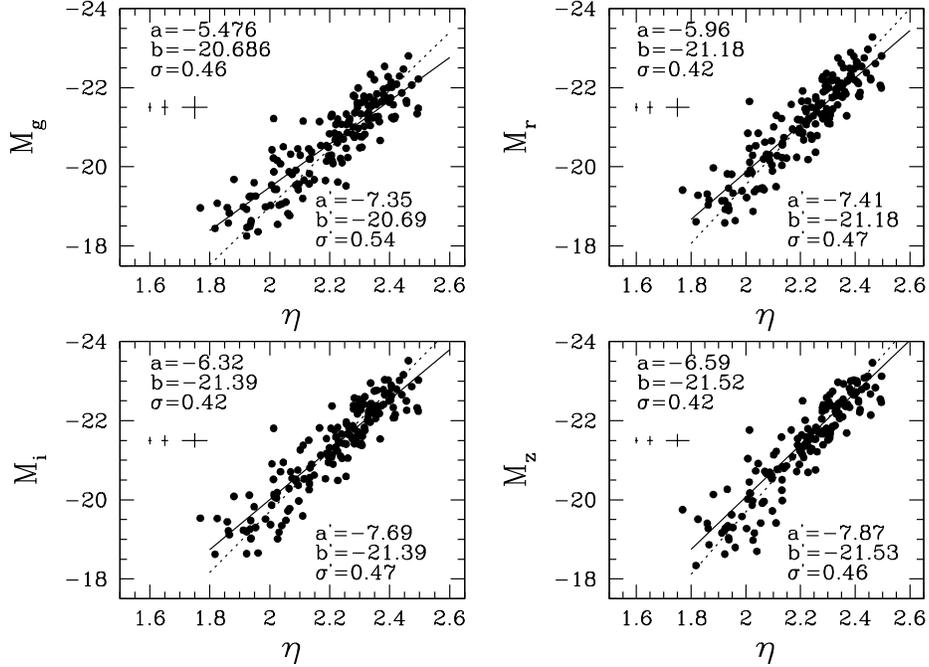}
\caption{TF relations for the full (\flagabc) 
sample in the $g$, $r$, $i$, and $z$-bands.  The
format is similar to Figure~\ref{fig:ibandTF}, except that we 
show the 10-th percentile, median, 90-th percentile 
1$\sigma$ errors crosses in each panel instead 
of plotting errors on each data point. All intercepts
$(b,b')$ are quoted at the $\eta_0$ values listed in Table 4.}
\label{fig:griz}
\end{figure}

\clearpage

\clearpage
	
\clearpage

\clearpage

\clearpage

\clearpage

\clearpage
\begin{figure}
\plotone{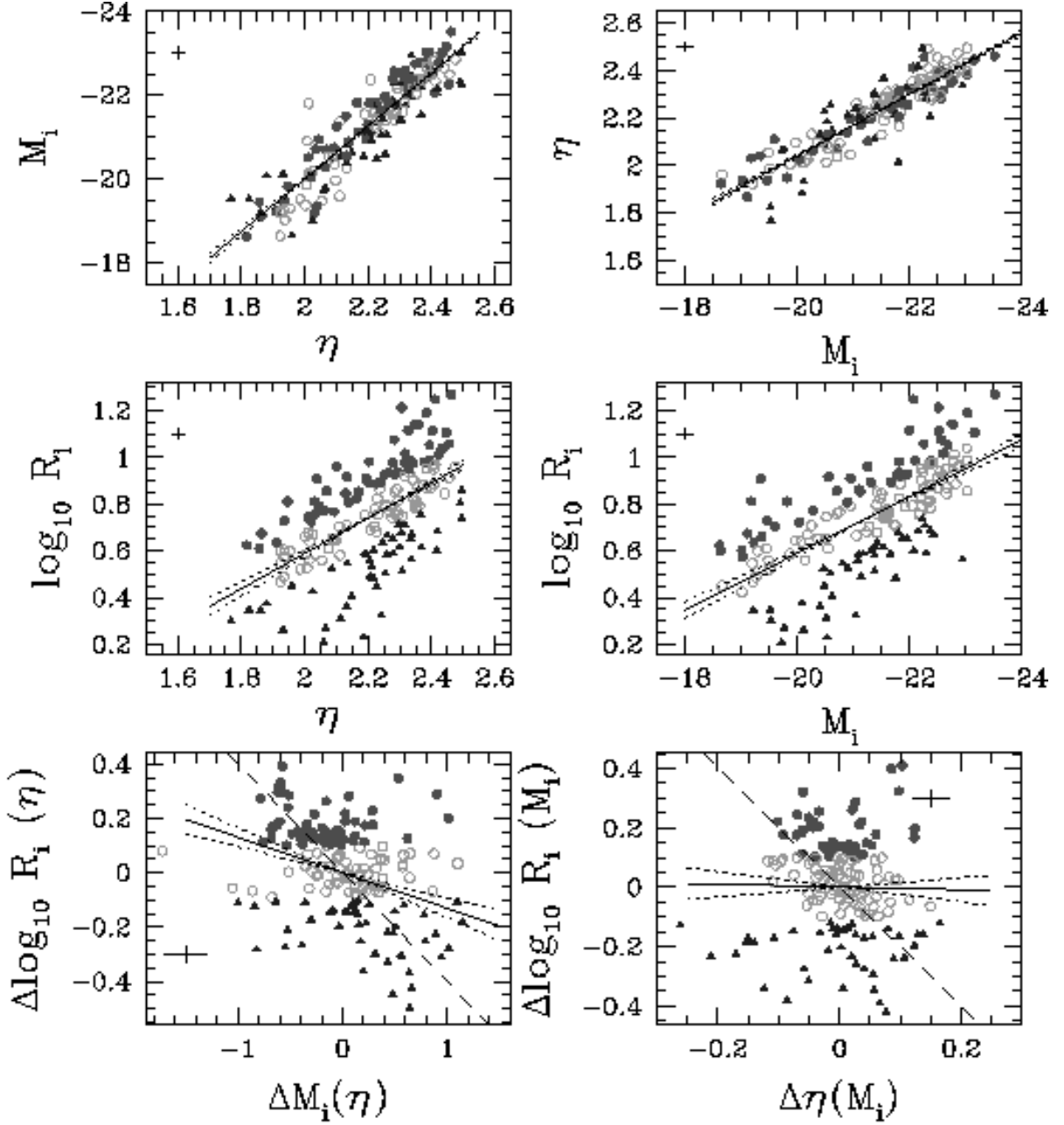}
\caption{Correlation of $i$-band TF residualfrom the mean relation 
between $i$-band half-light radius $R_i$ and $M_i$ or $\eta$.  The 
Format is otherwise the same as Figure 19.
}
\label{fig: radres}
\end{figure}

\clearpage

\clearpage

\clearpage
\begin{deluxetable}{lcc}
\tablewidth{0pt}
\tablecaption{Telescope and Spectrograph Parameters\label{tbl:obs}}
\tablehead{
\colhead{Parameter}  &
\colhead{Calar Alto } &
\colhead{MDM } }
\startdata
Telescope & 3.5 meter & 2.4 meter \\
Spectrograph & TWIN & CCDS \\
Grating & 1200 lines/mm & 600 lines/mm \\
Blaze & 7100 (1st order) \AA & 5875 (2nd order) \AA \\
Spectral Coverage & 6200-7300 \AA & 6500-6994 \AA \\
Slit Width & $1.5" \times 4'$ & $2" \times 5'$ \\
Spatial Scale & $0.56"/$pix & $0.41"$/pix \\
Dispersion & 0.55 \AA/pix & 0.41 \AA/pix \\
Read Noise & 3.9 e- & 7e- \\
Gain & 0.865  e-/ADU & 2.1 e-/ADU \\
Detector & SITe 2000 x 800 CCD  & Loral 1200x800 CCD \\
Net Instrument Eff. \tablenotemark{a} & 0.2 & $\sim$ 0.15 \\
\enddata
\tablenotetext{a}{telescope+spectrograph+detector}
\label{tbl:spectrographs}
\end{deluxetable}

\clearpage
\begin{deluxetable}{cccccccccc}
\tabletypesize{\scriptsize}
\rotate
\tablecaption{Photometric Parameters\label{tbl:phot}}
\tablewidth{0pt}
\tablehead{
\colhead{SDSS name} & 
\colhead{$d_{\rm lg}/{\rm Mpc}$} &
\colhead{$M_g$} &
\colhead{$M_r$} & 
\colhead{$M_i$} & 
\colhead{$M_z$} & 
\colhead{$g$-$r$} & 
\colhead{$D/T$} &
\colhead{$R_{i}$/kpc} &
\colhead{$b/a_i$} 
}
\startdata
J001006.62-002609.6\tablenotemark{a} &  143.75(4.39) &  -21.16(0.18) &  -21.30(0.15) &  -21.38(0.14) &  -21.22(0.09) &  0.49(0.03) &  0.92& 3.82(0.17) & 0.30(0.01) \\
J001708.77-005728.9 &  84.81(4.35) &  -19.57(0.15) &  -20.04(0.14) &  -20.26(0.13) &  -20.24(0.12) &  0.48(0.03) &  1.00& 5.11(0.28) & 0.58(0.01) \\
J001947.33+003526.7 &  79.65(4.34) &  -19.24(0.13) &  -19.82(0.12) &  -20.12(0.12) &  -20.27(0.12) &  0.54(0.01) &  0.54& 1.83(0.10) & 0.64(0.01) \\
J002025.78+004934.9 &  76.13(4.34) &  -20.29(0.21) &  -21.01(0.19) &  -21.26(0.17) &  -21.14(0.15) &  0.74(0.02) &  0.92& 9.15(0.55) & 0.41(0.01) \\
J003112.09-002426.4 &  87.00(4.35) &  -20.16(0.14) &  -20.87(0.14) &  -21.14(0.12) &  -21.33(0.12) &  0.74(0.01) &  0.79& 3.11(0.16) & 0.52(0.01) \\
J003629.24-004342.8\tablenotemark{b} & 152.58(4.40) & -19.98(0.06) & -20.78(0.06) & -21.18(0.06) & -21.47(0.06) & \nodata & \nodata & \nodata & \nodata \\
J003856.59+001757.3\tablenotemark{b} & 115.60(4.37) & -19.13(0.08) & -19.94(0.08) & -20.34(0.08) & -20.65(0.08) & \nodata & \nodata & \nodata & \nodata \\
J004239.34+001638.7 & 195.48(4.43) &  -21.82(0.16) &  -22.23(0.13) &  -22.41(0.11) &  -22.34(0.08) &  0.60(0.03) &  0.93& 9.45(0.40) & 0.55(0.01) \\
\enddata
\tablecomments{The Local Group barycenter distances are in Mpc, for
$H_0$ $=$ 70$\hubunits$. Absolute magnitudes and $g-r$ colors
are corrected for internal extinction, and K-corrected including the luminosity distance.  
The absolute magnitudes are calculated from the SDSS Petrosian magnitudes, 
colors are calculated from the SDSS model colors.  
Bulge-to-disk ratios, half-light radii, and and disk axis
ratios are computed from the $i$-band GALFIT models.  Table 2 is published in its entirety in the electronic 
edition of the Astronomical Journal. A portion is shown here for guidance regarding its form and content.}
\tablenotetext{a}{GALFIT magnitude used instead of SDSS Petrosian, 
because of poor pipeline estimation of the total observed magnitude.}
\tablenotetext{b}{\flagd\ galaxies with absolute magnitudes
and model colors that do not contain internal extinction corrections 
and K-corrections.}
\label{tbl:photparams}
\end{deluxetable}

\clearpage
\begin{deluxetable}{ccccccc}
\tabletypesize{\scriptsize}
\tablecaption{Arc-tangent Functional Fit and Rotation Speed\label{tbl:vel}}
\tablewidth{0pt}
\tablehead{
\colhead{SDSS name} & 
\colhead{$\Vcirc/\kms$} &
\colhead{$r_t$/arc-sec} & 
\colhead{$\Veight/\kms$} & 
\colhead{$\Vend/\kms$} &
\colhead{$V_{2.2}$/$\kms$} &
\colhead{flag}
}
\startdata
J001006.62-002609.6 & 160.82 &  2.79 & 129.27( 2.75) & 138.57( 2.40) & 125.90( 2.59) & 3 \\
J001708.77-005728.9 & 126.94 &  3.45 & 135.45( 6.08) & 125.61( 3.33) & 130.68( 5.41) & 3 \\
J001947.33+003526.7 &  71.67 &  0.84 &  85.51(10.59) &  80.07( 5.04) &  85.00(10.42) & 2 \\
J002025.78+004934.9 & 123.30 &  2.16 & 126.44( 7.67) & 121.88( 6.43) & 125.30( 7.51) & 3 \\
J003112.09-002426.4 & 150.47 &  1.13 & 161.36( 5.46) & 154.90( 3.56) & 159.90( 5.30) & 3 \\
J003629.24-004342.8 & \nodata & \nodata & \nodata & \nodata & \nodata & 4 \\
J003856.59+001757.3 & \nodata & \nodata & \nodata & \nodata & \nodata & 4 \\
J004239.34+001638.7 & 224.16 &  1.51 & 239.59( 3.83) & 240.77( 3.15) & 236.16( 3.61) & 1 \\
\enddata
\tablecomments{Parameters of
arc-tangent functional fits after 10 $\kms$ 
has been added to the velocity measurement errors on each data point.  
The velocities have been corrected for inclination and cosmological 
broadening.  Uncertainties 
are listed in parenthesis.  \flagd\ galaxies were observed but 
did not have enough \halpha\ to determine a rotation curve.  Table 3 is published in its entirety 
in the electronic edition of the Astronomical Journal. A portion is shown here for 
guidance regarding its form and content.
}
\tablenotetext{a}{Rotation curve data not used because of a poorly determined
SDSS pipeline isophotal position angle.}
\tablenotetext{b}{Rotation curve data not used because of a poorly determined
SDSS pipeline isophotal axis ratio.}
\end{deluxetable}

\clearpage
\begin{deluxetable}{ccccccccc}
\tabletypesize{\scriptsize}
\tablecaption{TF Parameters\label{tbl:tf}}
\tablewidth{0pt}
\tablehead{
\colhead{$y$} & 
\colhead{$x$} &
 \colhead{$x_0$} &
\colhead{$a$($\pm$)} & 
\colhead{$a'$($\pm$)} &
\colhead{$b$($\pm$)} & 
\colhead{$b'$($\pm$)} & 
\colhead{$\sigma$($\pm$)} & 
\colhead{$\sigma'$($\pm$)}}
\startdata
\cutinhead{$\Veight$ \flagabc\ N =162} \\
$M_g$ & ${\rm log} \Veight$ &   2.220 &$-5.476(0.226)$ & \nodata & $ -20.686(0.041)$ & \nodata & 0.457(0.034) & \nodata \\
${\rm log} \Veight$ & $M_g$ & $-20.607$ &$-0.136(0.006)$ & -7.35(0.32) &   2.209(0.006) & -20.687(0.049) &  0.073(0.006) & 0.536(0.044) \\
 $M_r$&  ${\rm log} \Veight$ &   2.220 & $-5.962(0.204$) & \nodata &  $-21.182(0.035$) & \nodata &  0.418(0.036) & \nodata \\
${\rm log} \Veight$ & $M_r$ & $-21.107$ & $-0.135(0.006)$ & -7.41(0.32) &   2.210(0.006) & -21.181(0.044) &  0.063(0.005) & 0.466(0.037) \\
$M_i$ &  ${\rm log} \Veight$ &   2.220 & $-6.321(0.220$) & \nodata & $-21.390(0.035)$ & \nodata &  0.423(0.035) & \nodata \\
 ${\rm log} \Veight$& $M_i$ & $-21.327$ & $-0.130(0.005)$ & -7.69(0.29) &  2.212(0.005) & -21.388(0.038) & 0.061(0.005) & 0.469(0.038) \\
$M_z$ &  ${\rm log} \Veight$ &   2.232 & $-6.590(0.238)$ & \nodata & $-21.518(0.034)$ & \nodata & 0.415(0.038) & \nodata \\
${\rm log} \Veight$ & $M_z$ & $-21.400$ & $-0.127(0.005)$ & -7.87(0.31) & 2.216(0.005) & -21.525(0.39) & 0.058(0.006) & 0.457(0.047) \\
\enddata
\tablecomments{Bivariate relations are fit with 
the model $y$=$a$($x$-$x_0$)+$b$ with a Gaussian intrinsic 
scatter of $y$ at fixed $x$, with dispersion $\sigma$.  Errors
listed for $a$, $b$, and $\sigma$ are computed from 100 
bootstrap trials, and the value of $x_0$ is chosen so that 
errors in $a$ and $b$ are uncorrelated.  $a'$, $b'$, and $\sigma'$ 
are the forward equivalent of the inverse relations.  }
\end{deluxetable}

\clearpage
\begin{deluxetable}{cccccc}
\tabletypesize{\scriptsize}
\tablecaption{Bivariate Relations and Residual Correlations\label{tbl:res}}
\tablewidth{0pt}
\tablehead{
\colhead{$y$} & 
\colhead{$x$} &
 \colhead{$x_0$} &
\colhead{a($\pm$)} & 
\colhead{b($\pm$)} & 
\colhead{$\sigma$($\pm$)} }
\startdata
$g$-$r$ & ${\rm log} \Veight$ &   2.220 &  0.576(0.055) & 0.552(0.0079) & 0.108(0.006) \\
$\Delta g-r$ ($\Veight$) & $\Delta   M_g$ ($\Veight$) &   0.00 &  0.067(0.018) &    0.000(0.0075) &   0.102(0.0067) \\
$g-r$ & $M_g$ & -20.607 & -0.063(0.018) &    0.544(0.0069) &   0.128(0.0062) \\
$\Delta   g-r$ ($M_g$) & $\Delta   {\rm log} \Veight$ ($M_g$) &  0.00 &  0.806(0.122) & 0.007(0.007) & 0.091(0.0069) \\
$\Delta   g-r$ ($\Veight$) & $\Delta  M_i$ ($\Veight$) & 0.00 & -0.003(0.021) & 0.000(0.008) & 0.108(0.006) \\
$g-r$ & $M_i$ & -21.327 & -0.075(0.008) &  0.547(0.009) &  0.115(0.008) \\ 
$\Delta   g-r$ ($M_i$) & $\Delta   {\rm log} \Veight$ ($M_i$) & 0.00 &  0.534(0.151) & 0.000(0.008) & 0.108(0.007) \\
${\rm log} R_i$ & ${\rm log} \Veight$ & 2.220 &  0.758(0.074) & 0.755(0.0115) & 0.169(0.010) \\
$\Delta   {\rm log} R_i$ & $\Delta   M_i$ & 0.00 & -0.130(0.036) & 0.000(0.011) & 0.155(0.0097) \\
${\rm log} R_i$ & $M_i$ &  -21.327 & -0.121(0.011) &  0.749(0.011) & 0.157(0.009) \\
$\Delta {\rm log} R_i$ ($M_i$) & $\Delta {\rm log} \Veight$ ($M_i$) & 0.00 & -0.046(0.195) & 0.00(0.011) & 0.157(0.009) \\
\enddata
\tablecomments{For modeling procedure, see note to 
Table~\ref{tbl:tf}.  }
\end{deluxetable}

\begin{deluxetable}{cccccc}
\tabletypesize{\scriptsize}
\tablecaption{ Alternative TF Fits \label{tbl:appx}}
\tablewidth{0pt}
\tablehead{
\colhead{$y$} & 
\colhead{$x$} &
 \colhead{$x_0$} &
\colhead{$a$($\pm$)} & 
\colhead{$b$($\pm$)} & 
\colhead{$\sigma$($\pm$)}}
\startdata
\cutinhead{$\Veight$\, N = 162 Weighted by \flagabc\ $\Mr$ distribution} \\
$M_g$ & ${\rm log} \Veight$ & 2.22 &  $-5.492(0.272)$ &  $-20.671(0.042)$ &  0.457(0.041)  \\
${\rm log} \Veight$ & $M_g$ & -20.607 & $-0.136(0.007)$ &  2.204(0.007) &  0.073(0.007) \\
$M_r$ & ${\rm log} \Veight$ & 2.22 &  $-5.977(0.268)$ &  $-21.166(0.0392)$ &  0.418(0.037)  \\
${\rm log} \Veight$ & $M_r$ &-21.107 &  $-0.135(0.006)$ &  2.208(0.00610) &  0.064(0.007) \\
$M_i$ & ${\rm log} \Veight$ & 2.22 &  $-6.334(0.286)$ &  $-21.347(0.039)$ &  0.423(0.038)  \\
${\rm log} \Veight$ & $M_i$ & -21.327 &  $-0.130(0.005)$  & 2.211(0.006) &  0.061(0.006)  \\
$M_z$ & ${\rm log} \Veight$ & 2.22 &  $-6.607(0.260)$  & $-21.496(0.044)$  & 0.415(0.041)  \\
\enddata
\tablecomments{Same as Table~\ref{tbl:tf}, but with different
weighting of data points, internal extinction correction,
velocity definitions, or sample selection, as indicated.
}
\end{deluxetable}

\clearpage
\begin{deluxetable}{ccccc}
\tabletypesize{\scriptsize}
\tablecaption{Tully-Fisher Relations in the Literature\label{tbl:TFotherworks}}
\tablewidth{0pt}
\tablehead{
\colhead{study} & 
\colhead{data} &
 \colhead{bias corrected} &
\colhead{slope} & 
\colhead{scatter} }
\startdata
PT88\tablenotemark{a} & $R$-HI & no & -7.64(0.29) & $<$ 0.25 mag \\
PT92\tablenotemark{b} &  $R$-HI & no & -8.23 & small \\
W96\tablenotemark{c} &   $r$-H$\alpha$  &  yes  & -8.195(0.263)  & 0.3  \\ 
V01\tablenotemark{d} & $R$-HI & no & -9.6(0.4) & 0.15 \\
K02\tablenotemark{e} & $R$-band-H$\alpha$ & no & -9.70(0.27) & 0.42 \\
PT88\tablenotemark{a} & $I$-HI & no & -7.85(0.29) & $<$ 0.25 mag \\
PT92\tablenotemark{b} &  $I$-HI & no & -8.72 & small \\
W95\tablenotemark{f} &  $I$-HI & no & -7.865(0.162) &  0.392 \\
\enddata
\tablecomments{Tully-Fisher relations from different samples.  The 
intercepts are not listed due to different assumptions for $H_0$, 
and different values for the assumed internal extinction corrections.  
The scatter is the intrinsic scatter.}
\tablenotetext{a}{\cite{pt88} study of spirals in Virgo and Ursa Major, having no morphological peculiarities, depth of the cluster included in analysis.}
\tablenotetext{b}{\cite{pt92} study of galaxies with Cephied/RR-Lyrae, and PNe distances. Sample contains 15 galaxies.}
\tablenotetext{c}{\cite{wil96} analysis of the CF sample, with cz $\geq$ 3000 $\kms$ in the Local Group frame, with the total scatter.}
\tablenotetext{d}{V01 forward TF relation for the RC/FD Sample (excluding NGC 3992), using $V_{flat}$, and
the most likely intrinsic scatter (depth of cluster assumed to be 0.17 mag).}
\tablenotetext{e}{K02 inverse TF relation for galaxies with $M_R$ $<$ $-18$, weighted inverse slope.}
\tablenotetext{f}{\cite{wil95} TF relations for spiral galaxies in clusters.}
\tablenotetext{g}{\cite{gio97} cluster TF relation for 782 spirals.}
\tablenotetext{h}{C97 analysis of CF data with optical rotation curves.}
\tablenotetext{i}{C97 analysis of MAT sample having optical and HI data.}
\end{deluxetable}

\end{document}